\documentclass{aa}  

\usepackage{graphicx}
\usepackage{txfonts}
\usepackage{lipsum}
\usepackage{subcaption}         
\usepackage{lscape}             
\usepackage{placeins}           

\usepackage{xcolor}
\usepackage{mathtools}
\usepackage{bbm}
\usepackage[normalem]{ulem}
\usepackage{url}
\usepackage{amsmath}
\usepackage{multirow}
\usepackage[backref=page]{hyperref}
\usepackage{fdsymbol}
\usepackage{cancel}

\newcommand{\gaia}{{\it Gaia}}
\newcommand{\Gaia}{{\it Gaia}}
\newcommand{\gdr}[1]{\gaia~DR#1\xspace}

\newcommand{\bprp}{\ensuremath{G_{\rm BP}-G_{\rm RP}}\xspace}
\newcommand{\Ha}{\ensuremath{\rm H\alpha}\xspace}
\newcommand{\Hb}{\ensuremath{\rm H\beta}\xspace}
\newcommand{\Hg}{\ensuremath{\rm H\gamma}\xspace}
\newcommand{\SNRHa}{\ensuremath{\left(\textrm{S/N}\right)_{\rm H\alpha}}\xspace}
\newcommand{\SNRHb}{\ensuremath{\left(\textrm{S/N}\right)_{\rm H\beta}}\xspace}
\newcommand{\SNRHg}{\ensuremath{\left(\textrm{S/N}\right)_{\rm H\gamma}}\xspace}
\newcommand{\orcit}[1]{\protect\href{https://orcid.org/#1}{\protect\includegraphics[width=8pt]{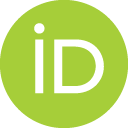}}}

\begin{document}

   \title{XP-TEAL: \Gaia\ XP tool for emission and absorption lines}
   \titlerunning{XP-TEAL: \Gaia\ XP tool for emission and absorption lines}
   \authorrunning{J.~M.~Carrasco et al.}

   \subtitle{I. Balmer lines in open clusters}

   \author{J.~M. Carrasco\orcit{0000-0002-3029-5853}\inst{1,2,3}\corrauth{carrasco@fqa.ub.edu}  
   M. Weiler\orcit{0000-0002-3007-3927}\inst{1,2,3} \and
   S. Malhotra\orcit{0000-0002-5509-0168}\inst{1,2,3} \and  
   F. Anders\orcit{0000-0003-4524-9363}\inst{1,2,3} \and
   C. Jordi\orcit{0000-0001-5495-9602}\inst{3,4} \and
   E. Masana\orcit{0000-0002-4819-329X}\inst{1,2,3}\and
   J. Casta\~{n}eda\orcit{0000-0001-7820-946X}\inst{5,2,3} \and
   F. Boix\inst{1,2,3}    
        }

   \institute{Departament de F\'{\i}sica Qu\`antica i Astrof\'{\i}sica (FQA), Universitat de Barcelona (UB), Mart\'{\i} i Franqu\`es 1, E-08028 Barcelona, Spain
   \and
   Institut de Ci\`encies del Cosmos (ICCUB), Universitat de Barcelona (UB), Mart\'{\i} i Franqu\`es 1, E-08028 Barcelona, Spain
   \and
   Institut d'Estudis Espacials de Catalunya (IEEC), Edifici RDIT, Campus UPC, 08860 Castelldefels (Barcelona), Spain
   \and   
   Institut d'Estudis Catalans, c. del Carme, 47, E-08001 Barcelona, Spain
   \and
   DAPCOM Data Services, c. dels Vilabella, 5-7, 80500 Vic,
Barcelona, Spain.
}

   \date{Received March 17, 2026}

  \abstract
   {The third {\Gaia} data release ({\gdr3}) includes blue and red low-resolution spectrophotometry for more than 200 million sources. These {\Gaia} XP spectra are represented by sets of coefficients multiplying linear combinations of basis functions related to Hermite functions. In a previous work we introduced a method that maximally exploits the information encoded in the {\Gaia} spectrophotometric coefficients in order to identify absorption and emission lines.}
   {Our aim is to detect hydrogen Balmer emission and absorption lines in stars belonging to open clusters included in the released {\gdr3} blue and red spectrophotometry, and to characterise the Balmer line strengths as a function of colour.} 
   {We refined our XP tool for emission and absorption lines (XP-TEAL), using derivatives of the internal spectral coefficients to analyse hydrogen Balmer lines in the {\gdr3} XP spectra. We tested our results using stars in open clusters, considering both absorption and emission features.}
   {The analysis of the hydrogen lines in the spectra of open cluster stars allows us to assess the performance and limitations of the method. We find an empirical dependence of the absorption-line equivalent widths on the intrinsic colour of the stars, which allows us to derive precise colour excesses caused by interstellar extinction. We also identify 1256 stars with emission lines, 356 of which have not previously been reported in the literature.}
   {We successfully tested our XP-TEAL code on stars in clusters and have made both the source code and the online version available to the community. Despite the low resolution of {\gaia} spectrophotometry, our code is able to extract useful line information.}


   \maketitle
\nolinenumbers

\section{Introduction}
\label{sec:introduction}

The third \gaia\ data release ({\gdr3}) \citep{Vallenari2023, Prusti2016} provides, for the first time, low-resolution blue (BP) and red (RP) spectrophotometry for approximately 219 million sources \citep{DeAngeli2023}. Most of these sources have apparent $G$ magnitudes lower than 17.65~mag, but the catalogue also includes some fainter objects, such as white dwarfs and quasars \citep{DeAngeli2023}. The BP spectral range extends from 330~nm to 680~nm, while the RP spectral range extends from 640~nm to 1050~nm \citep{Carrasco2021}. Both spectrophotometers have resolving powers below $R=\lambda/\Delta\lambda=100$, depending on wavelength \citep{Carrasco2021}. In the following, we use the term XP when referring to BP or RP spectra without distinction. 

The XP spectra are described in {\gdr3} as linear combinations of Hermite functions \citep{DeAngeli2023}. From the same set of coefficients, the spectra can be reconstructed either in an internally calibrated form (still convolved with the total instrument response) or as externally calibrated spectra (in physical flux and wavelength units). The internal spectra express the flux in photo-electrons per second, per unit pseudo-wavelength (a pixel-based coordinate), within the {\Gaia} aperture, whereas the external spectra combine BP and RP into a single spectrum in energy units per unit time, wavelength, and detector area. The construction of the externally calibrated spectra involves deconvolving the low-resolution spectra with the line spread function (LSF) \citep{Montegriffo2023}.

Since the internally calibrated spectra contain complete spectral line information, we used them and avoided the external calibration step, which introduces additional sources of uncertainty. We exploited the specific representation of the internally calibrated XP spectra in {\gdr3} as linear combinations of Hermite functions, following the formalism developed by \citet{weiler2023}. We briefly summarise this formalism in Sect.~\ref{sec:method}. Using this formalism we developed a web tool that is also available for use by the reader (XP-TEAL, Appendix~\ref{sec:xpteal}.).

In this work we use the {\gaia} low-resolution XP spectra to study the Balmer hydrogen lines of stars belonging to open clusters (OCs) in {\gdr3}. We derive equivalent widths (EWs) for three Balmer lines, \Ha, \Hb, and \Hg\ in OC stars (Sect.~\ref{sec:clusters}). In addition, we analyse the sources with detected emission lines in OCs (Sect.~\ref{sec:emissionOCs}). We estimate colour excesses for OCs by deriving an empirical dependence of absorption-line EWs on the stars' intrinsic colour (Sect.~\ref{sec:colourexcessOCs}). Finally, in Sect.~\ref{sec:conclusions} we summarise our main results and conclusions.

\section{Methodology}
\label{sec:method}

{\gdr3} XP spectra are described by a set of coefficients and a set of Hermite basis functions \citep{Carrasco2021,DeAngeli2023}. The same set of coefficients can be combined with different bases to sample either internal or external spectra; these are related through a transformation (shift and scale) to match the respective XP pseudo-wavelength grids (see Fig.~4 in \citealt{Carrasco2021}). In {\gdr3}, these Hermite bases are further orthogonally transformed to reduce the number of significant coefficients \citep{DeAngeli2023}.

\begin{figure*}[!htbp]
    \centering
\includegraphics[width=0.4\textwidth]{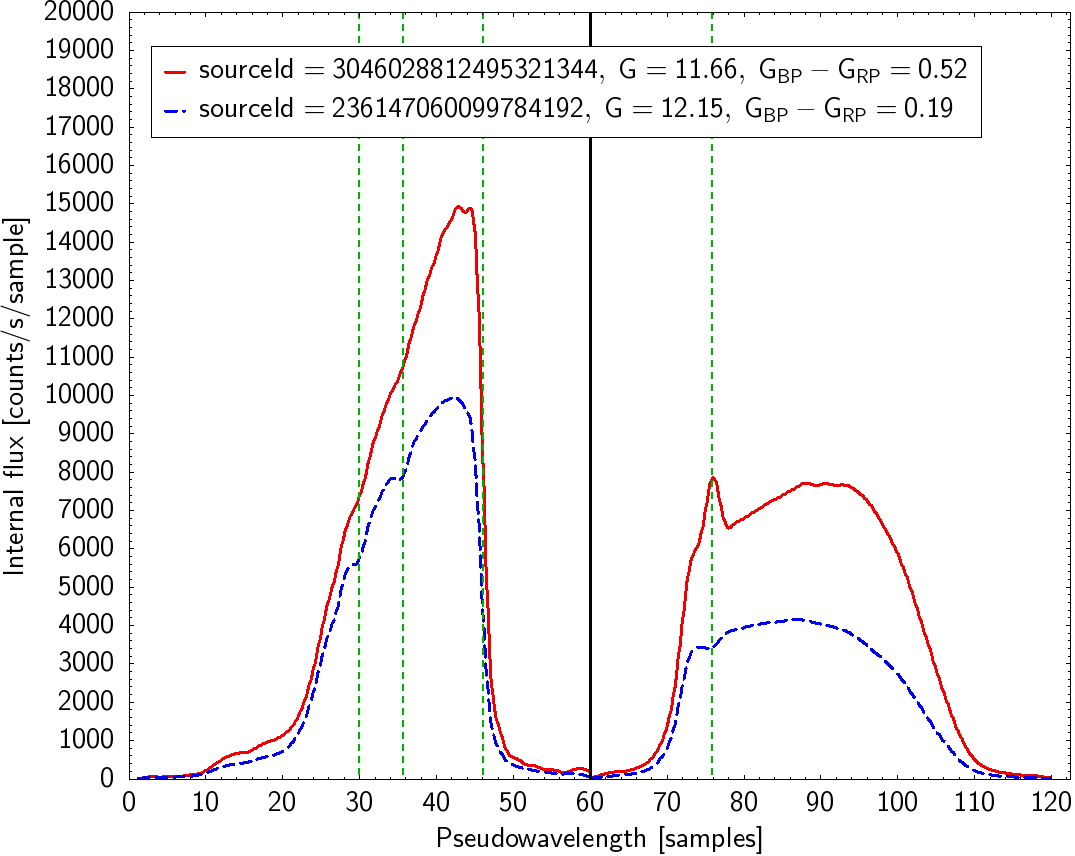}    
\includegraphics[width=0.4\textwidth]{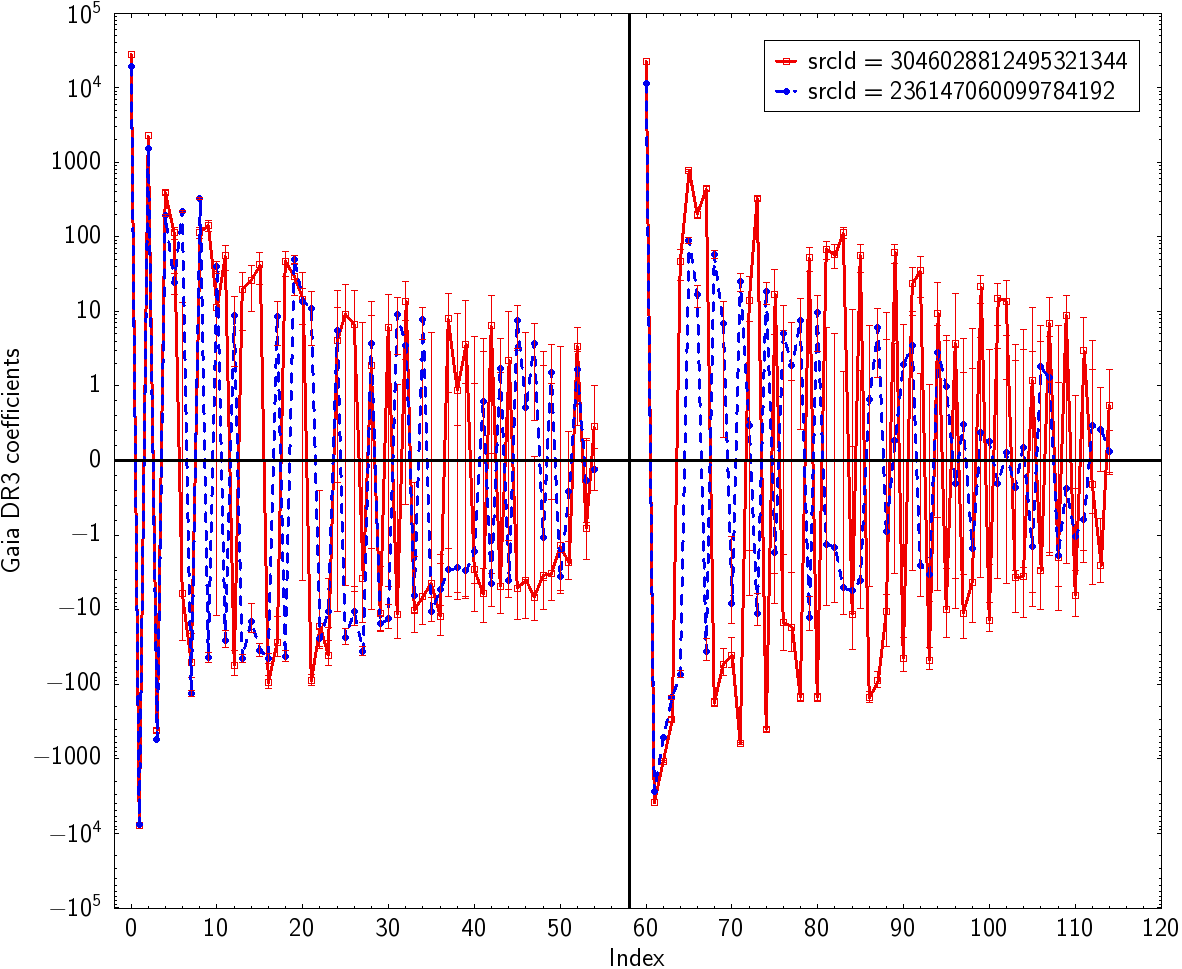}
    \caption{
    Example \gdr3\ internal XP spectra (left) and their coefficients (right) for two sources: one with \Ha\ in emission (solid red line) and one with Balmer absorption lines (dashed blue line). The RP pseudo-wavelengths (left panel) and index coefficients (right panel) are shifted by +60 so that BP and RP appear side by side. The solid vertical black line separates the BP (left) and RP (right) spectra. Left panel: Pseudo-wavelengths represented in the abscissa increase from shorter (left) to longer (right) wavelengths;  dashed vertical green lines mark the positions of \Ha\ in RP and of \Ha, \Hb\ and \Hg\ in BP (from right to left) using the nominal \Gaia\ dispersion relation. Right: Ordinate uses a symmetric logarithmic scale to better display the small high-order coefficients relative to the large low-order ones.
\label{fig:internalspectra} 
    }   
\end{figure*}

Hermite functions have a particularly convenient property: the derivative of any linear combination of Hermite functions is itself a linear combination of Hermite functions. Thus, given the coefficient vector of an internal spectrum, the coefficients of any derivative can be obtained by multiplying the original coefficient vector by an appropriate matrix. Moreover, the roots of a linear combination of Hermite functions are the eigenvalues of a `non-standard companion matrix' associated with that combination. Standard eigenvalue techniques therefore allow us to efficiently compute all roots of a given combination. Together, these properties make it straightforward to locate local minima and maxima in the internally calibrated XP spectra. This framework was developed in detail by \citet{weiler2023} and implemented in the algorithm XP tool for emission and absorption lines (XP-TEAL\footnote{XP-TEAL web tool is available 
at \url{https://xpteal.fqa.ub.edu}, and the Python source code at \url{https://github.com/malhotrasagar15/xpy_teal}.}; see Appendix~\ref{sec:xpteal}), which we employ here under the narrow-line approximation (see Sect. 8 in \citealt{weiler2023}).

Spectral lines appear as local extrema in the internally calibrated spectra, or in their second derivatives when the lines are weak and the gradient of the underlying spectrum (continuum plus response curve) is large. Neglecting intrinsic line widths, the LSF (itself expressed as a linear combination of Hermite functions) can be scaled and added or subtracted until a smooth interpolation is obtained. Smoothness is defined as the disappearance of either the second or the fourth derivative, depending on whether the line manifests directly in the spectrum or in its second derivative. This procedure yields the line EW from an internally calibrated spectrum with a modest computational cost.

We adopted the formalism of \citet{weiler2023} to measure Balmer-line strengths using the XP-TEAL narrow-line mode. We analysed three Balmer lines: \Ha\ at 656.470\,nm (better observed in RP than in BP, because of the higher resolution at this wavelength), and \Hb\ at 486.274\,nm and \Hg\ at 434.173\,nm (both observed only in BP). We selected these lines because they have the highest signal-to-noise ratio (S/N) in \gaia\ XP spectra and are mostly free from blends \citep{Creevey2022,Fouesneau2023}. Line strengths are expressed as EWs. Following \citet{weiler2023}, we adopted the convention that negative EWs correspond to absorption and positive EWs to emission.

The Balmer lines of main-sequence stars are usually observed in absorption, with the line strength varying systematically with effective temperature (that is, spectral type). Hydrogen absorption peaks for A-type stars and decreases towards both earlier and later types \citep{Maury1897, Cannon1901}.

As illustrated for two example {\gdr3} sources in Fig.~\ref{fig:internalspectra}, the internal XP spectra clearly carry information on the hydrogen Balmer lines. The pseudo-wavelength positions of \Ha, \Hb, and \Hg, computed from the nominal dispersion law for {\gdr3}\footnote{Following \cite{weiler2023}, we also used an extra shift of +0.3 in pseudo-wavelength for {\Ha} in RP in left panel of Fig.\ref{fig:internalspectra} in order to better match the spectral line.}, are indicated by green dashed vertical lines. The \Ha\ line is located in the overlap region between BP and RP and is therefore accessible from both instruments. The \Hb\ and \Hg\ lines are only present in BP. The corresponding XP coefficient vectors and their errors are also shown in Fig.~\ref{fig:internalspectra}. The low-order coefficients, which mainly encode the continuum, are larger and have smaller errors, while higher-order coefficients, which carry most of the line information, are smaller and comparable to their errors.

In this work we computed the EWs for the Balmer lines for the sample of \gaia\ objects with XP spectra in OCs (see Sec.~\ref{sec:clusters}). For each line, we also derived the ratio between the measured EW and its uncertainty, which we refer to as the S/N of the line. 

\section{Application to OC stars}
\label{sec:clusters}

We investigated the hydrogen Balmer lines in stars that belong to OCs. Open clusters (OCs) are particularly useful for our purposes because their member stars share a common origin (age and metallicity), a coherent space motion (which helps establish membership), and a similar distance, so that apparent magnitudes can be used to construct meaningful colour–magnitude diagrams (CMDs) without using individual parallaxes. Since all members lie approximately along the same line of sight, they are also typically affected by a similar amount of interstellar reddening.

\begin{figure*}[!htbp]
     \centering
     \includegraphics[width=0.33\linewidth]{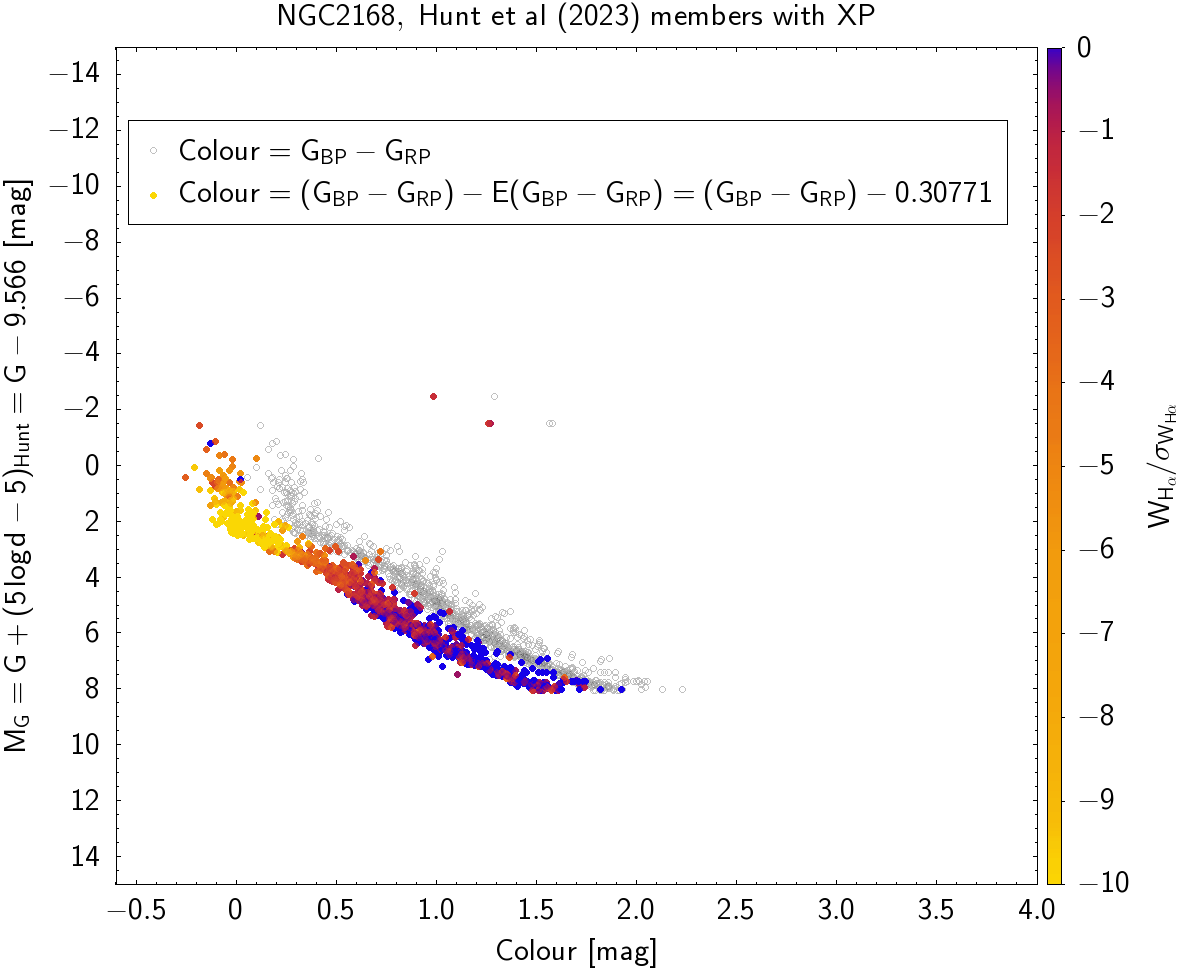}
     \includegraphics[width=0.33\linewidth]{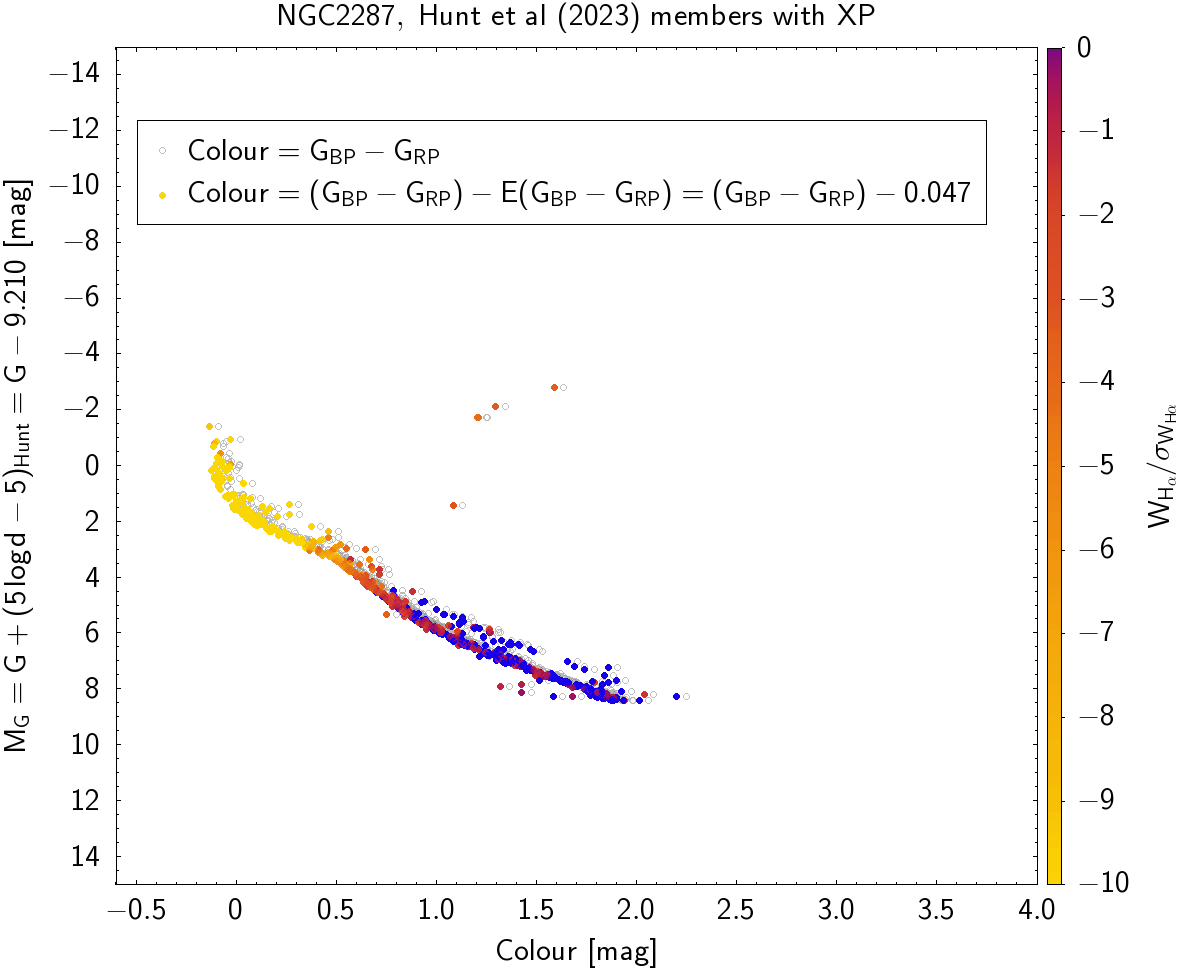}    
     \includegraphics[width=0.33\linewidth]{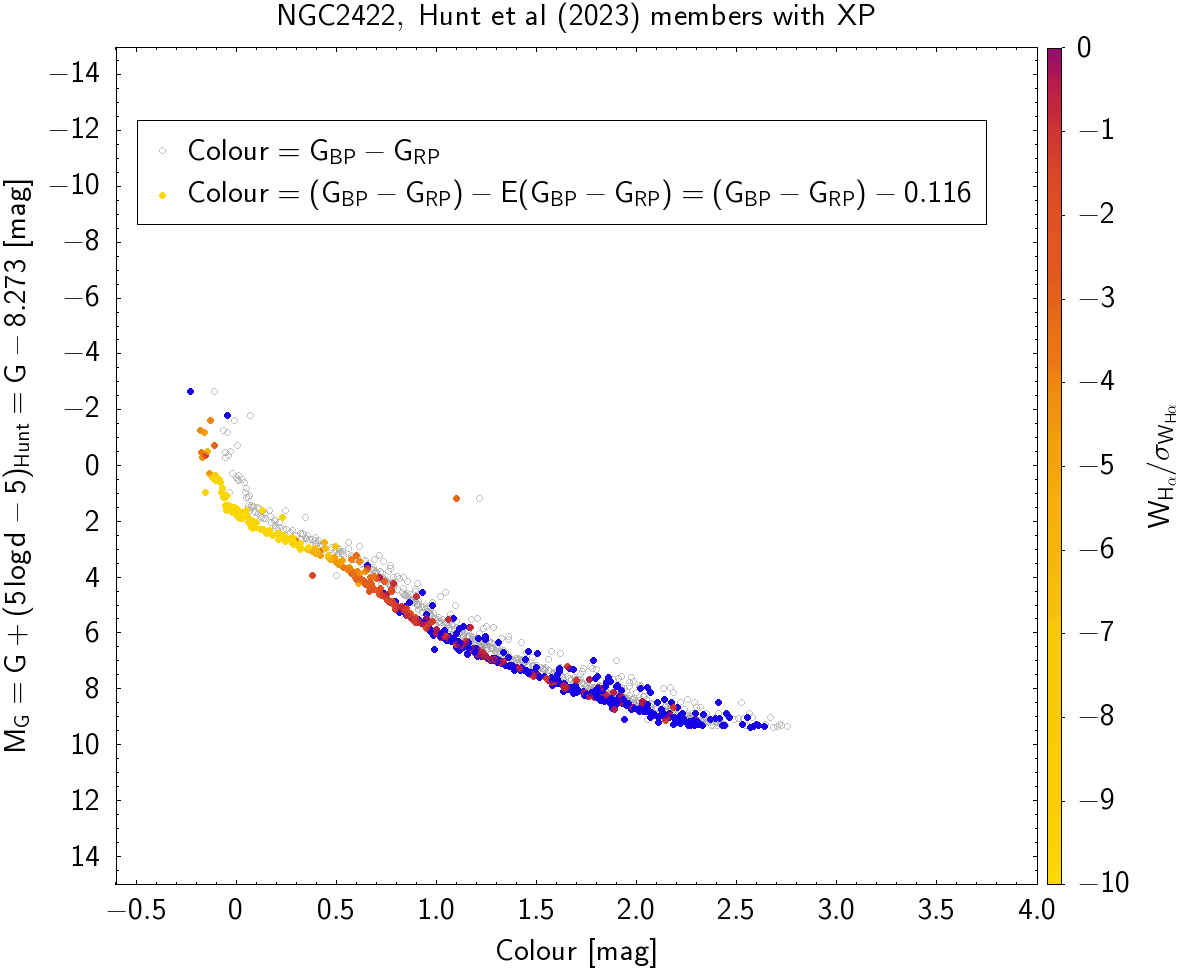}
     \includegraphics[width=0.33\linewidth]{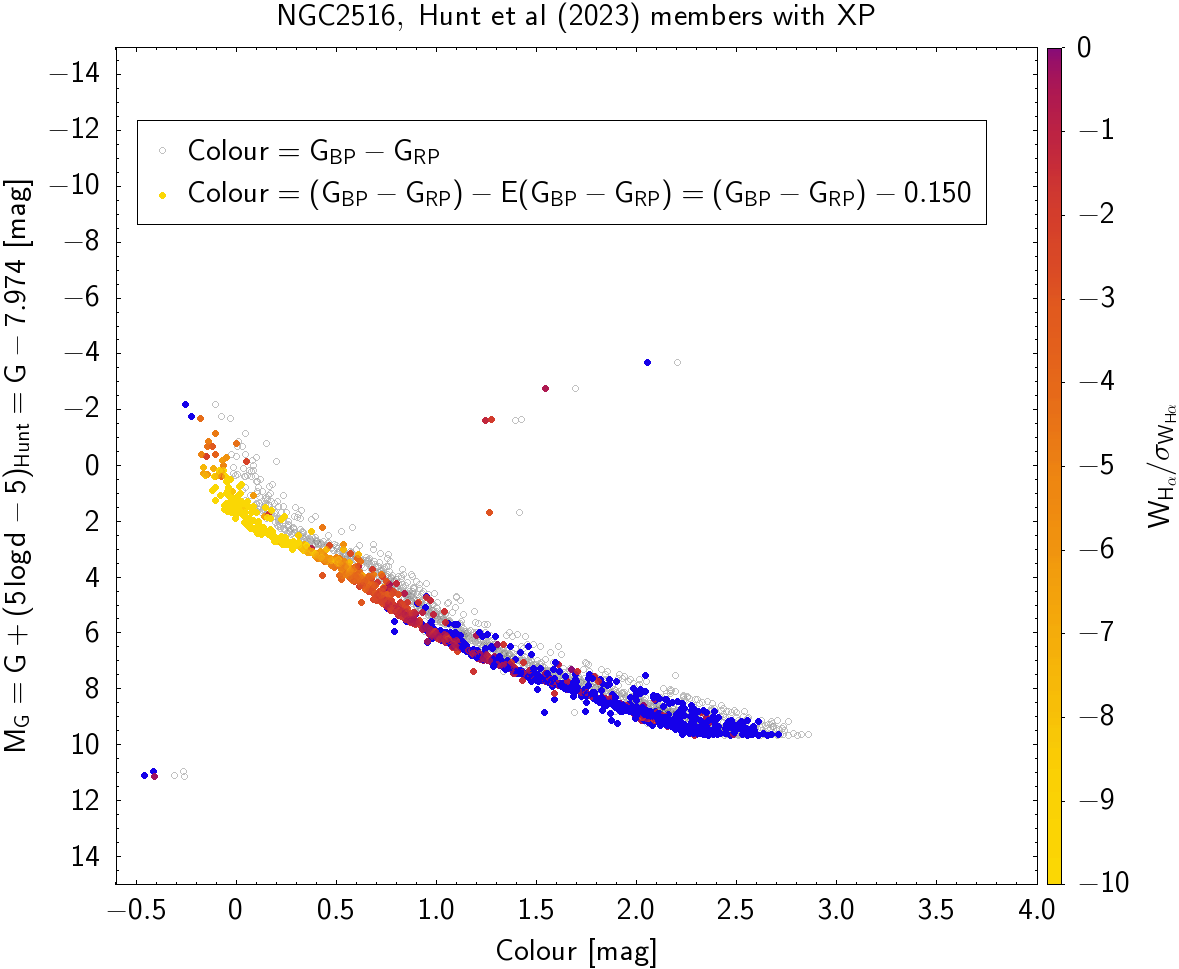}
     \includegraphics[width=0.33\linewidth]{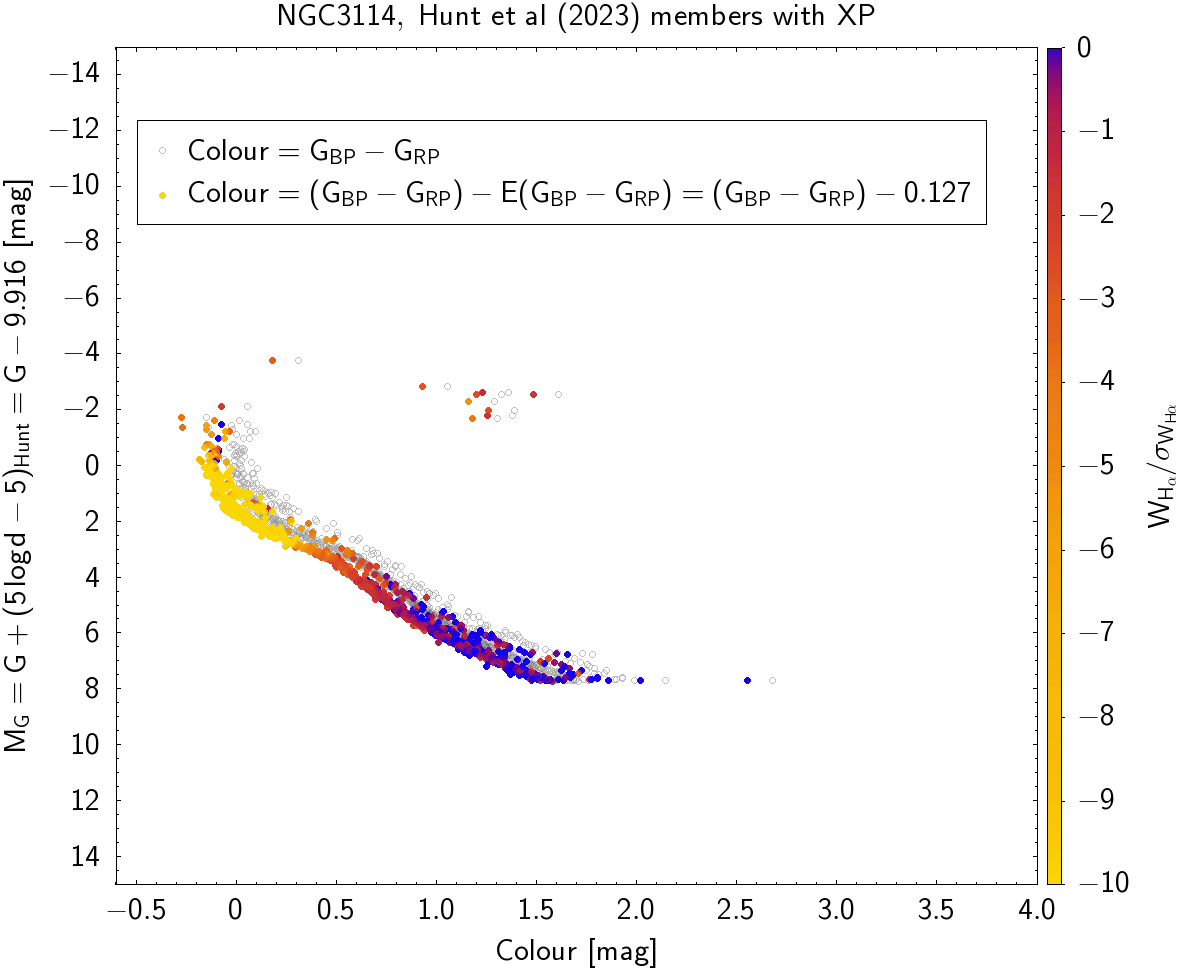}  
     \includegraphics[width=0.33\linewidth]{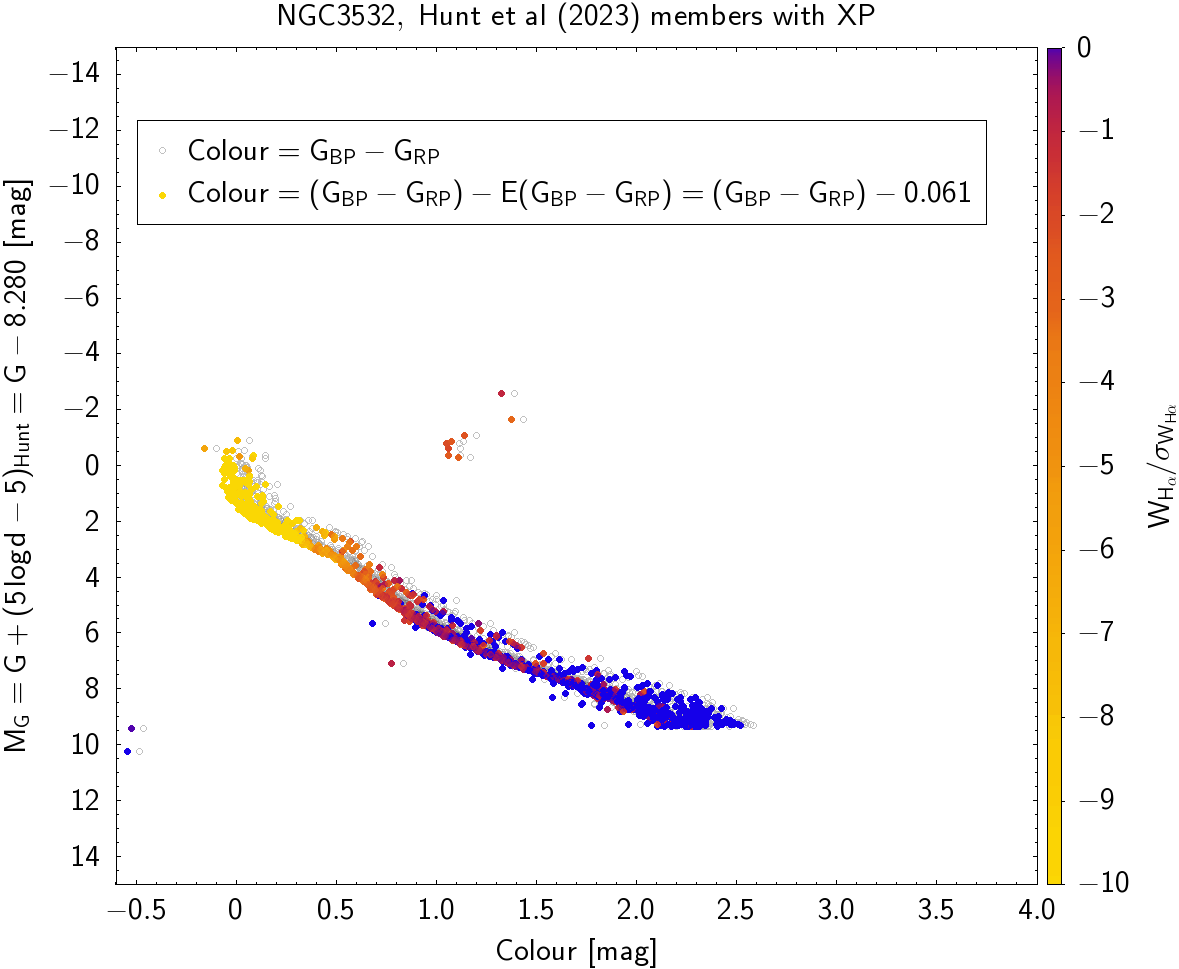}
     \caption{
     \gdr3 CAMDs for an illustrative set of young OCs from \citet{Hunt2023} with prominent blue main sequences. The vertical axis was scaled to absolute magnitudes using distance modulus values from \citealt{Hunt2023}. Grey points represent apparent \bprp\ colours, while coloured symbols show intrinsic colours on the x axis, corrected using the colour excesses derived in Sect.~\ref{sec:colourexcessOCs}. The colour scale encodes the \Ha\ EW obtained in this work.
     }
    \label{fig:cmd}
 \end{figure*}

Because of their relatively young ages (the age distribution peaks at $\log \tau [\rm yr] \sim8.2$; e.g. \citealt{Anders2021}), some OCs still host blue main-sequence stars, which are ideal for tracing the evolution of hydrogen Balmer lines with colour or temperature in the \gaia\ XP spectra. Figure~\ref{fig:cmd} presents several colour-absolute magnitude diagrams (CAMDs) for a selection of such young clusters for illustrative purposes.

Several studies \citep[e.g.]{Cantat2018,Cantat2020,Cantat2020Portrait, CastroGinard20, CastroGinard22, Hunt2023} provide cluster identifications and memberships from \gaia. In this paper we adopted the catalogue of \citet{Hunt2023} as our primary reference and used a total of 7135 cluster candidates (6785 of them classified as OCs; see Table~\ref{tab:clusters}). Of the 1\,291\,929 stars listed by \citet{Hunt2023}, 608\,014 sources have XP spectra in \gdr3, with 989 of these stars appearing as members of more than one cluster in \citet{Hunt2023}. The catalogue distinguishes OCs (kind = `o') from globular clusters (kind = `g'). As mentioned above, we were interested in OCs and therefore excluded the globular clusters from our analysis. We also excluded 19 additional globular clusters from \citet{Vasiliev2021}.

\begin{figure}
    \centering
    \includegraphics[width=0.9\columnwidth]{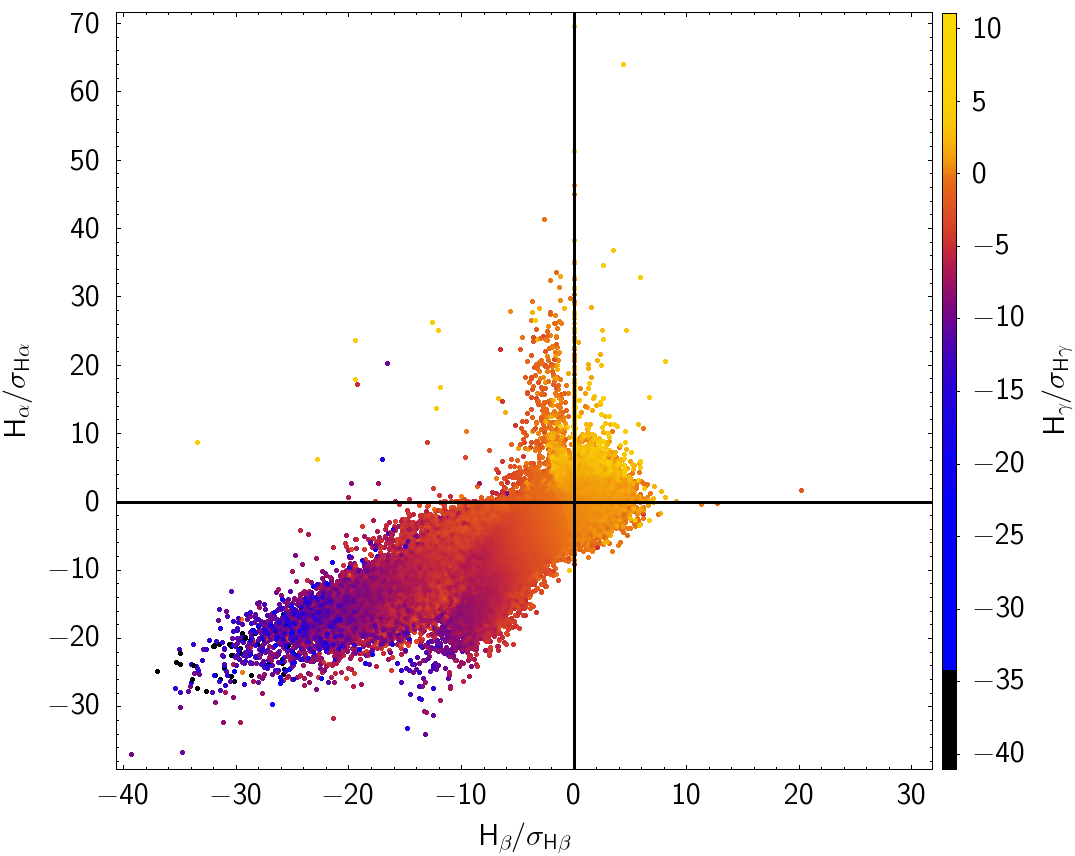}
    \includegraphics[width=0.9\columnwidth]{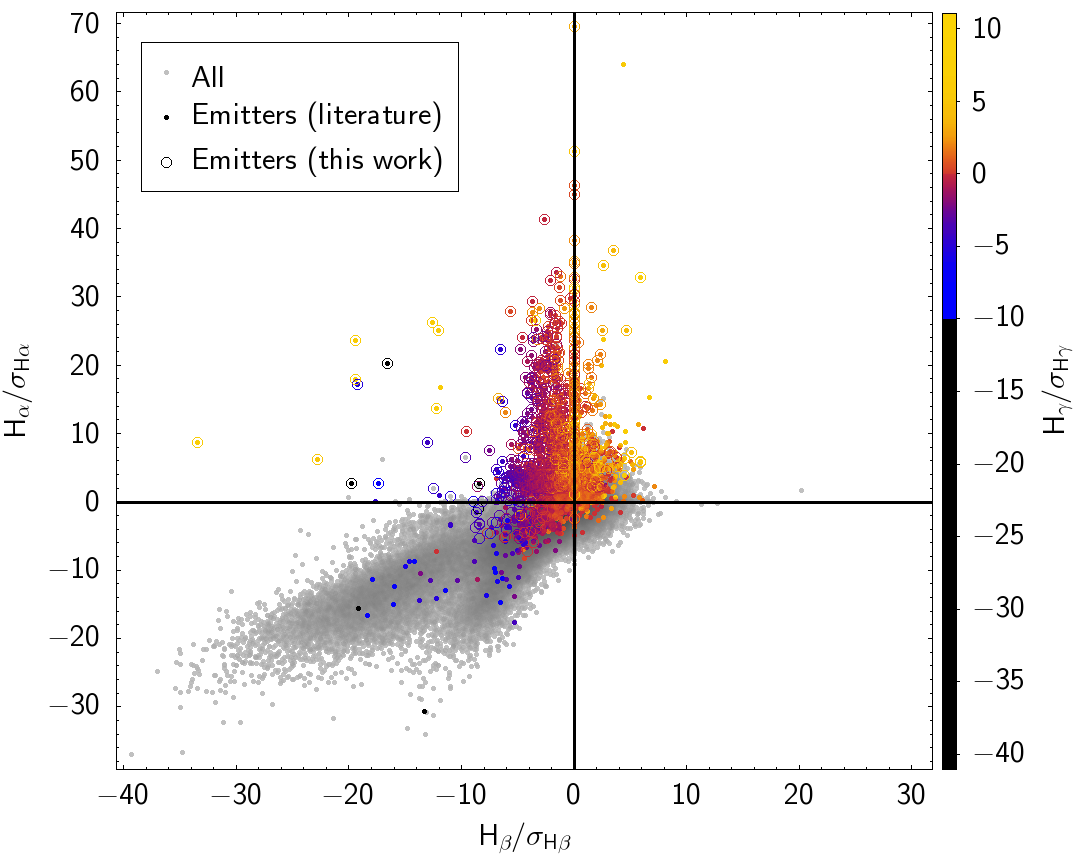}    
   \caption{
   Ratio of EW to its uncertainty (S/N) for \Ha\ (ordinate), \Hb\ (abscissa), and \Hg\ (colour scale) for all selected OC stars with XP spectra. Bottom panel: same but for known emission-line stars from the literature (solid symbols) and emission candidates identified in Sect~\ref{sec:emissionOCs} (open symbols).
   }
    \label{fig:SNRHaHbHg}
\end{figure}

Figure~\ref{fig:SNRHaHbHg} shows \SNRHa, \SNRHb and \SNRHg for all selected OC stars with XP spectra. Both panels are colour-coded by \SNRHg. The top panel highlights the (expected) correlation between the EWs of the three measured Balmer lines in absorption (with two distinct sub-branches in \Hb versus \Ha). The bottom panel highlights stars that show \Ha in emission (see Sec.~\ref{sec:emissionOCs}), showing that this emission is often (but not always) accompanied by \Hb and \Hg in absorption. This is also expected, since many different stellar sub-classes can show \Ha in emission.

\section{Emission lines in OC stars}
\label{sec:emissionOCs}

\begin{figure*}[!htbp]
    \centering
    \includegraphics[width=0.66\columnwidth]{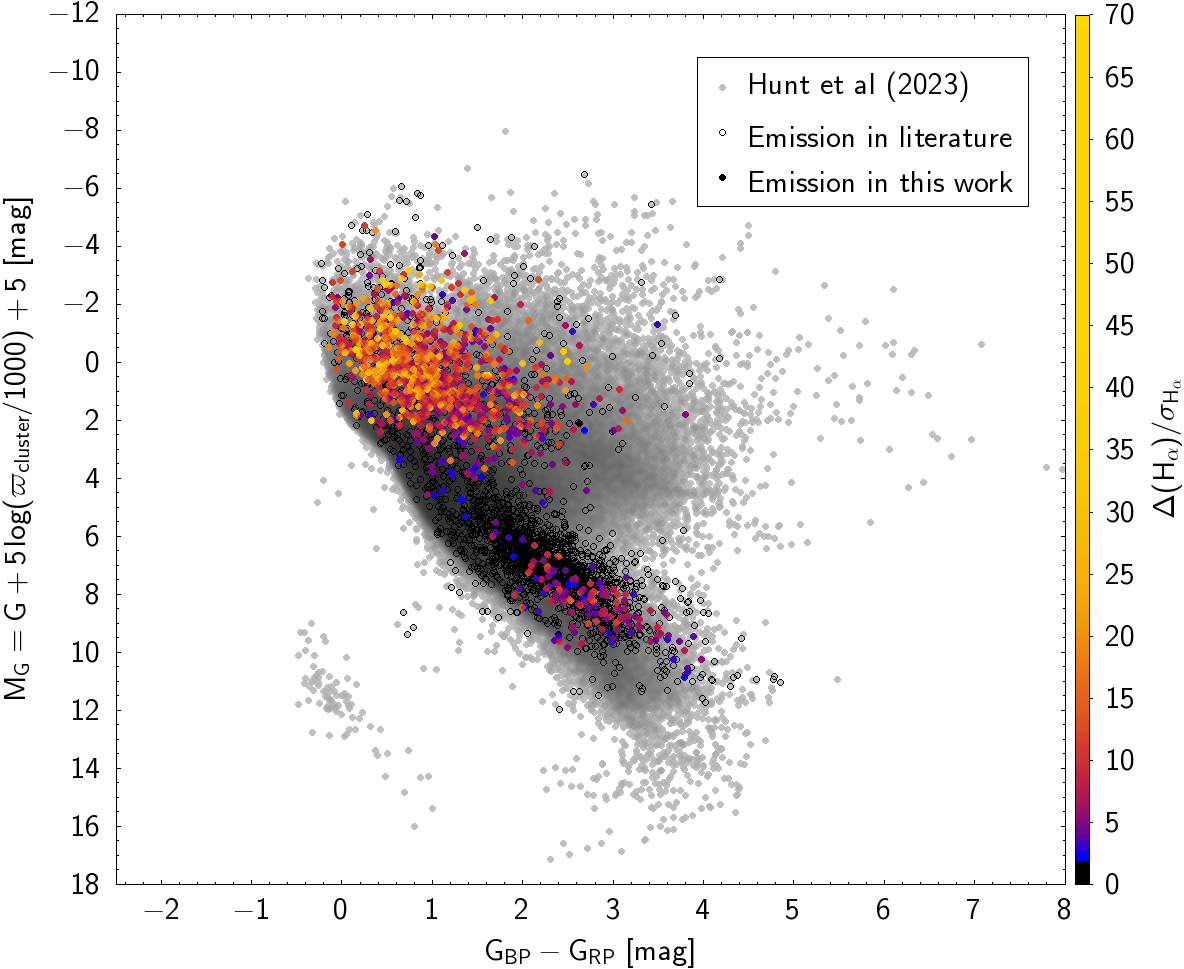}    
    \includegraphics[width=0.66\columnwidth]{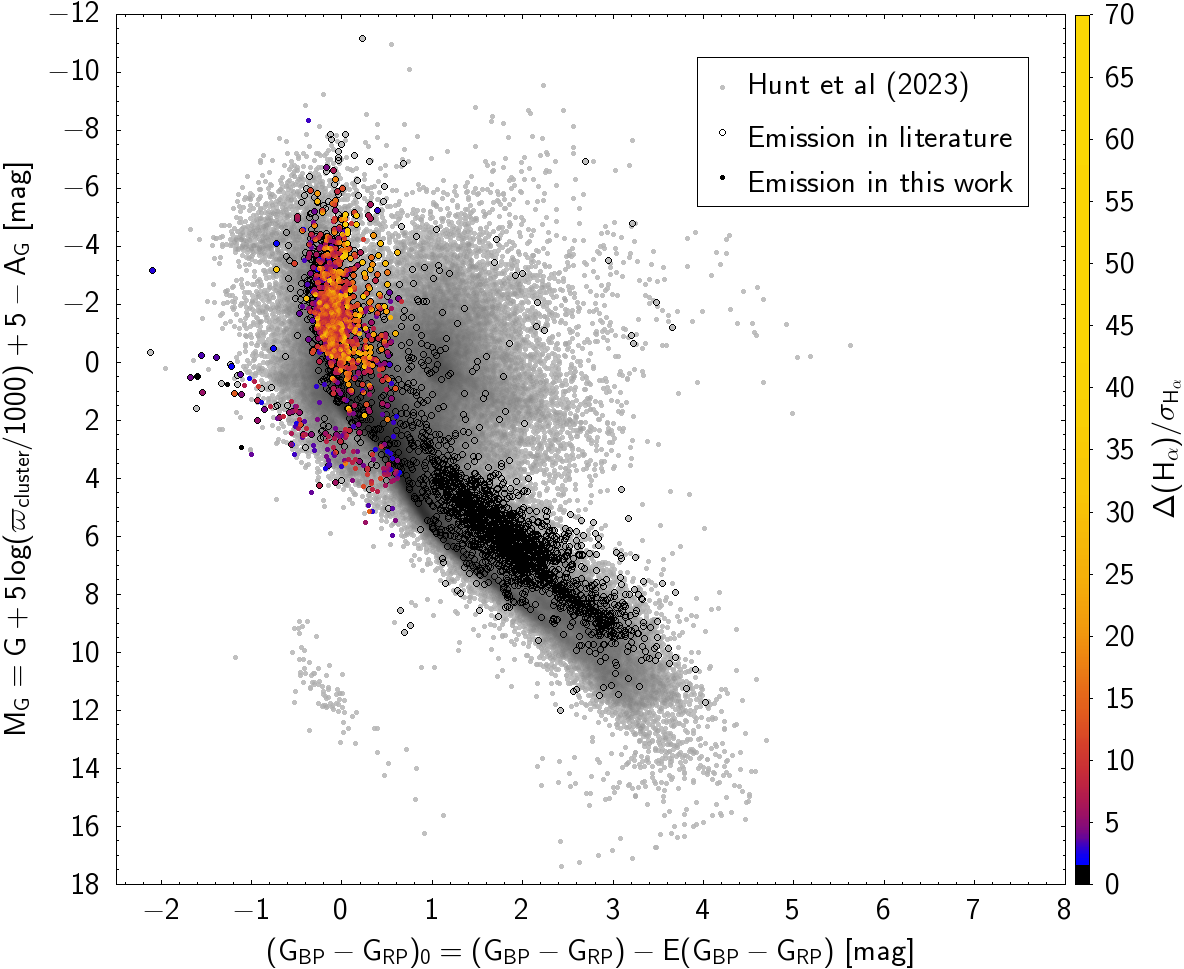}
    \includegraphics[width=0.66\columnwidth]{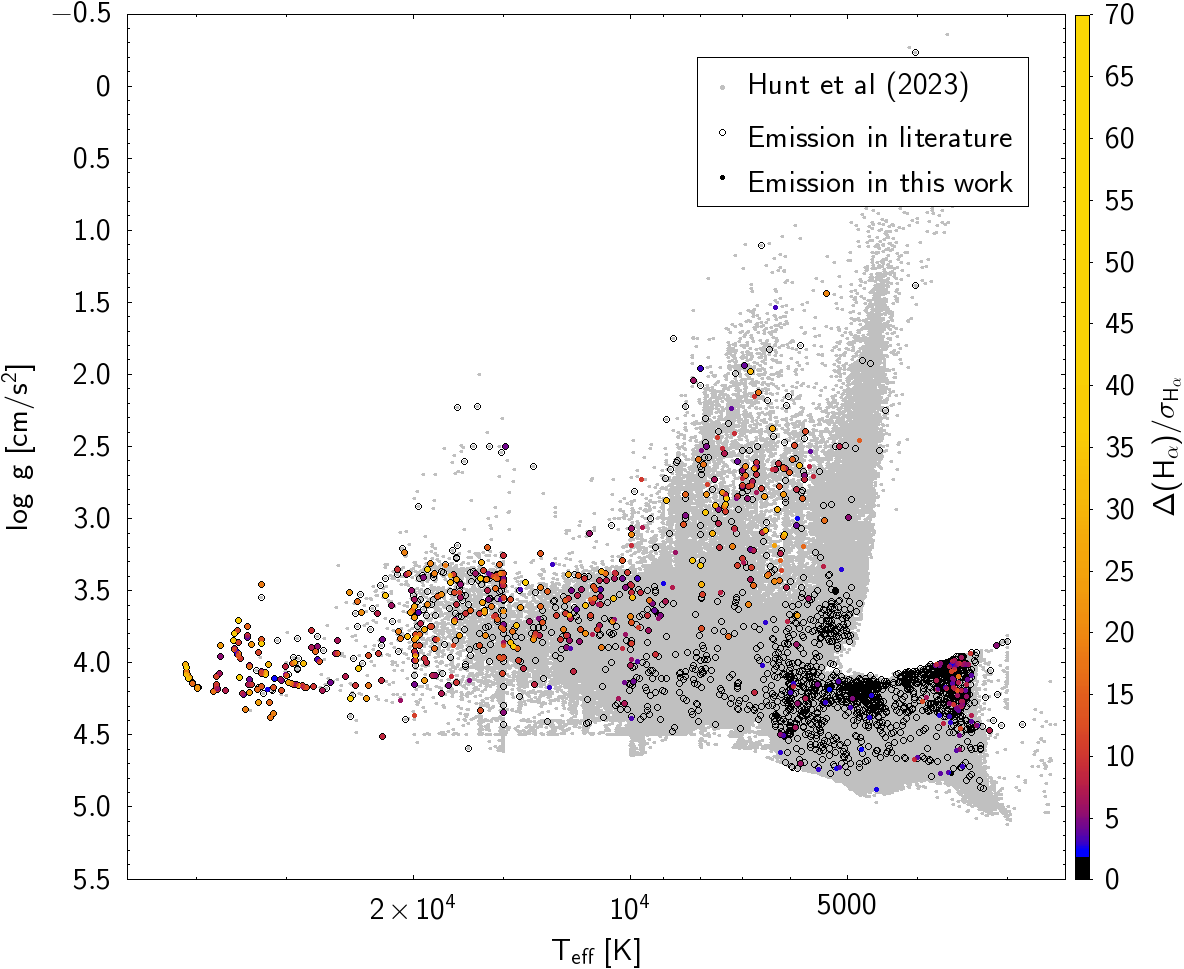}
    \caption{Location of emission-line cluster members in the Hertzsprung-Russell-like diagrams. Open symbols represent previously known emission stars; filled symbols indicate new emission candidates from Sect.~\ref{sec:emissionOCs}. Left: observed CAMD (using the parallax of the cluster from \citep{Hunt2023}).  
    Centre: De-reddened CAMD using \(A_G\) from BaSeL-3.1-based relations in Eq.~\ref{eq:AGvsEBPRP_Basel}. 
    Right: GSP-Phot \citep{Creevey2022,GSPPHOT} parameters derived for the same sources.
    \label{fig:HRdiagramEmitters}}
\end{figure*}

Among the 608\,159 stars from \citet{Hunt2023} with XP spectra analysed here, \gdr3 (in particular, the ESP-ELS pipeline)
flags 1400 sources as emission-line candidates. In addition, we find 2787 stars with emission-type classifications in SIMBAD astronomical database \citep{SIMBAD}, either in the `main\_type', `other\_types', or `sp\_type' fields. Combining both lists yields 3602 stars previously known or suspected of hosting emission lines (see Figs.~\ref{fig:SNRHaHbHg} lower panel and \ref{fig:HRdiagramEmitters}).

In this paper, we classify a star as an emission candidate if its \Ha\ EW is greater than three standard errors above zero. Using this criterion we identify 1\,256 emission candidates (1\,157 in OCs), which we list in Table~\ref{tab:members} and plot as coloured points in Fig.~\ref{fig:HRdiagramEmitters}. Of these, 900 (845 in OCs) stars were already classified as emitters in the literature. The remaining 356 (312 in OCs) candidates are new emission-line candidates that, to our knowledge, had not been reported before. 
Another paper with an in-depth analysis of the emission-line stars in the full {\gdr3} catalogue in preparation by Malhotra et al. (in prep).

Figure~\ref{fig:HRdiagramEmitters} shows how the \Ha\ emission varies in different regions of the CAMD and as a function of the effective temperature and surface gravity, as reported by the \gdr3\ GSP-Phot algorithm \citep{GSPPHOT}, which derives the astrophysical parameters from XP spectra. Emission sources populate the upper main sequence, the pre-main sequence, and regions expected for evolved hot stars, in line with their spectroscopic classifications. The emission strength depends on both the temperature and the apparent magnitude in a non-trivial manner. 
Within a given OC, these quantities are correlated because cooler stars are typically fainter. In a cluster-by-cluster comparison, distance and extinction also play an important role: for more distant and/or reddened clusters, the S/N of the Balmer lines is generally lower, as expected.

\begin{figure*}[!htbp]
    \centering
    \includegraphics[width=0.66\columnwidth]{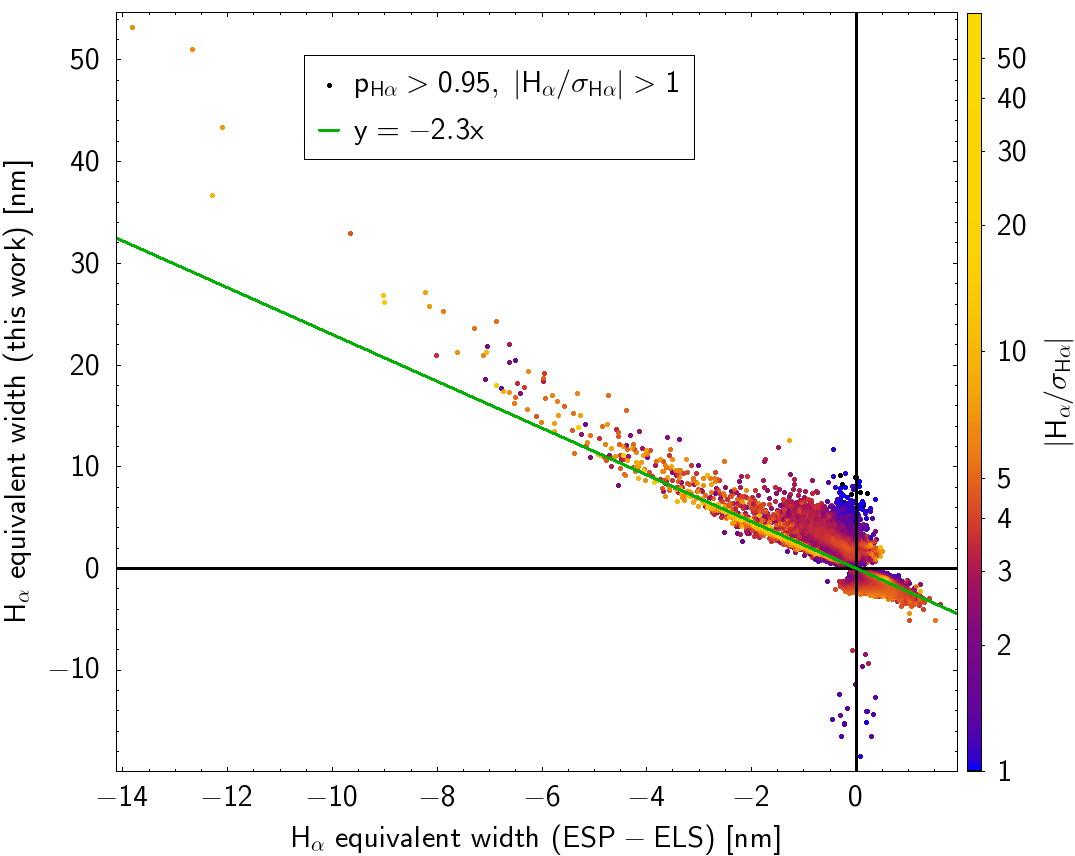}    
    \includegraphics[width=0.66\columnwidth]{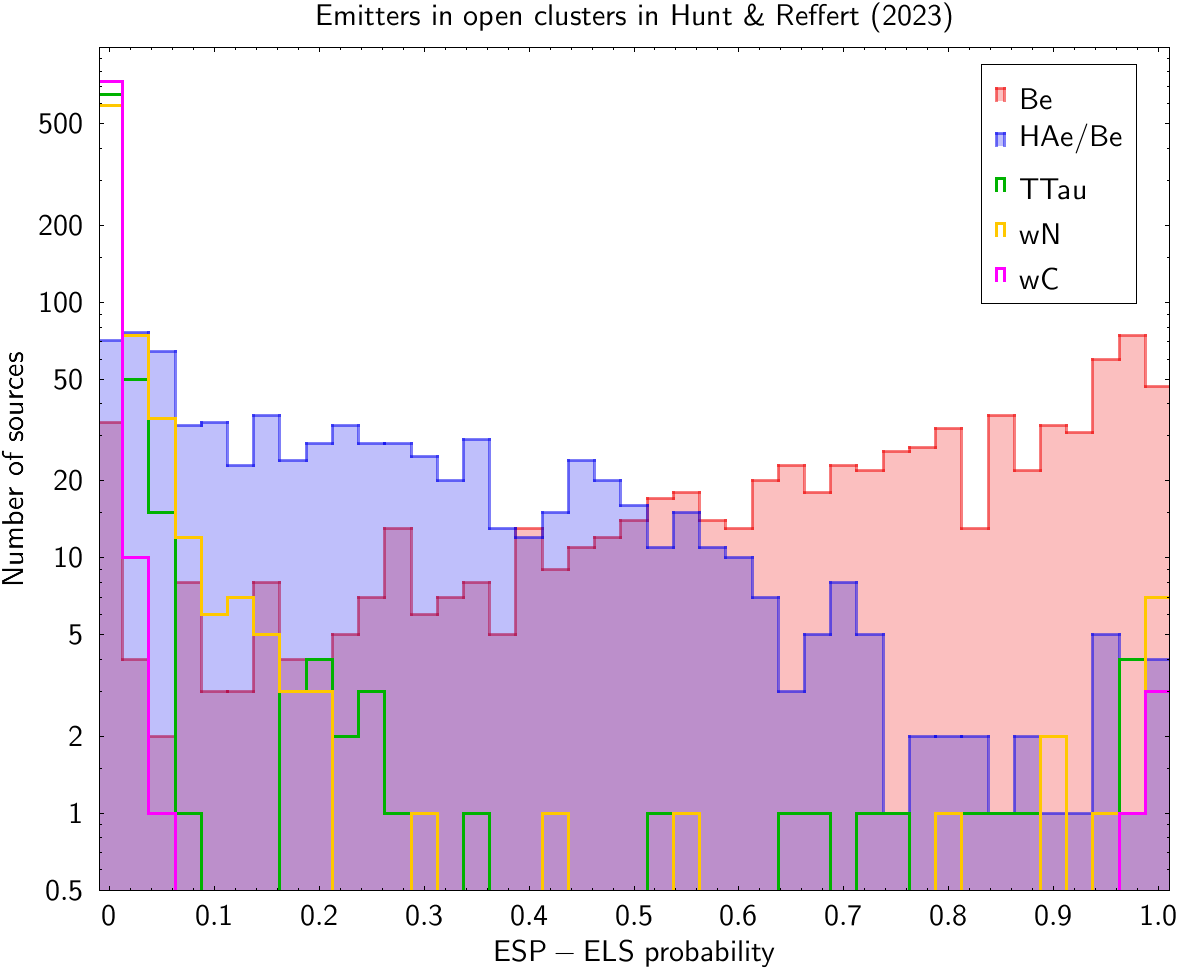}
    \includegraphics[width=0.66\columnwidth]{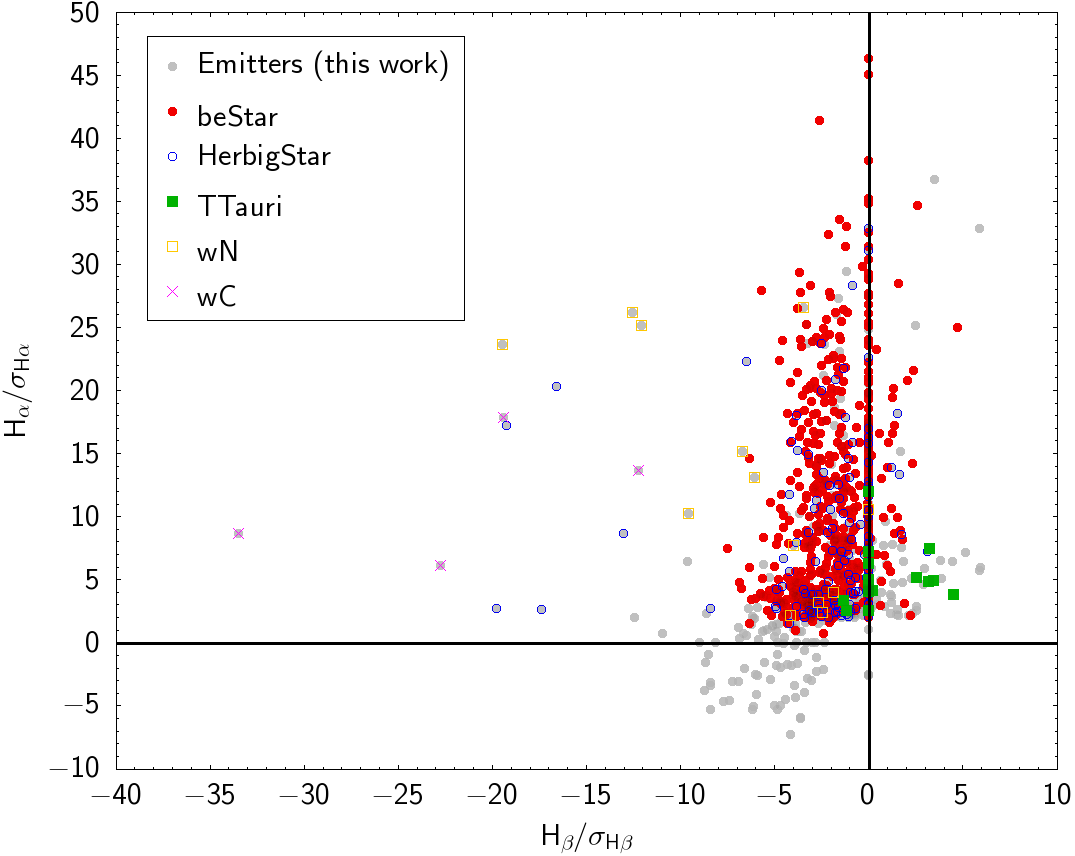}
    \caption{Comparison between \Ha\ EWs measured in this work and the \gdr3 ESP-ELS results \citep{Creevey2022}. Left: Direct EW comparison for OC members with \(|\SNRHa|>1\) and \(p_{\Ha}>0.95\), colour-coded by |\SNRHa|. The difference in sign convention (emission positive here and negative in \citealt{Creevey2022}) produces the anti-correlation. The green line indicates a slope of $-2.3$. 
    Middle: Probability distributions for different emission types (Be, Herbig Ae/Be, T~Tauri, carbon and nitrogen Wolf–Rayet) from \citet{Creevey2022}. Right: \SNRHa versus \SNRHb coloured by ESP-ELS types; grey points without colour coding represent new emission candidates from this work.}
    \label{fig:classELSemitters}
\end{figure*}

\label{sec:espels}

The Extended Stellar Parametrizer for Emission-Line Stars \citep[ESP-ELS;][]{Creevey2022, Fouesneau2023} provides independent \Ha\ EW estimates 
for emission-line stars in \gdr3, referred to as `pseudo-equivalent widths'. Although we also used XP spectra, we used the spectral coefficients directly, whereas ESP-ELS uses the sampled spectra derived from them to perform its analysis. We compared ESP-ELS results with our XP-TEAL-based measurements in Fig.~\ref{fig:classELSemitters}. We restricted the comparison in the left panel to sources with \(|\SNRHa|>1\) and a $p$-value \(p_{\Ha}>0.95\)\footnote{The closer the $p$-value is to unity, the higher the significance of the local extremum (see Appendix~\ref{sec:xpteal})}, so that both methods operate in a reasonably high S/N regime. The agreement between the two sets of measurements is good, although our absolute EWs tend to be about a factor of two larger. We also note that the sign convention is the opposite: we adopted positive values for emission, whereas ESP-ELS uses negative values, which leads to a negative correlation in the linear fit in the left panel of Fig.~\ref{fig:classELSemitters}. The green line in that panel indicates a slope of $-2.3$, which shows that the scale of EWs between this work and \gdr3 ESP-ELS differs by more than a factor of two. Other datasets also report differences of this order of magnitude with respect to the literature \citep{Shridharan2022}.

Figure~\ref{fig:classELSemitters} summarises the ESP-ELS classification of the emission candidates in OCs. Based on probabilistic classifications, ESP-ELS assigns 637 (587 in OCs) objects as Be stars, 131 (128 in OCs) as Herbig Ae/Be stars, 17 as T~Tauri stars, 13 as nitrogen-sequence Wolf–Rayet stars, and four as carbon-sequence Wolf–Rayet stars (all T~Tauri and Wolf–Rayet stars lie in OCs).

\begin{figure}
    \centering
    \includegraphics[width=0.9\linewidth]{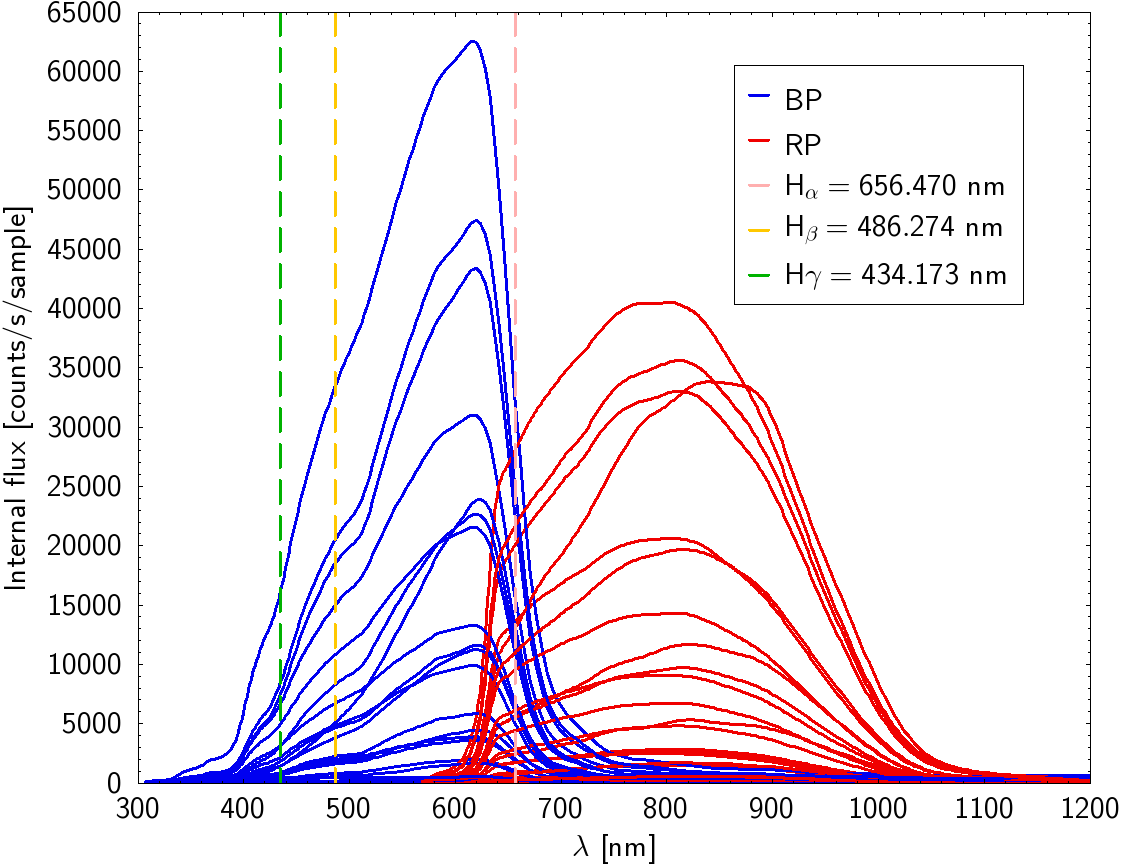}
    \caption{{\gdr3} 
    internally calibrated XP spectra for 45 sources classified as emitters in the literature but not identified by our methodology using XP spectrophotometry (yielding EW equal to zero in all three Balmer lines). Vertical dashed lines indicate the three Balmer lines analysed in this work (from right to left: \Ha, \Hb, and \Hg).}
    \label{fig:EmittersNotDetected}
\end{figure}

In total we find 3465 previously known emission stars in OCs when combining SIMBAD and ESP-ELS information. Our Balmer-based selection confirms 845 of these among our 1256 candidates. Our method does not flag the remaining 2620 OC stars as emission sources (see Figs.~\ref{fig:SNRHaHbHg} and \ref{fig:HRdiagramEmitters}). For 1511 of these sources our algorithm assigns an EW equal to zero for at least one Balmer line, meaning that no significant local extremum is detected in the spectrum or its second derivative at the expected line position. For the others, the \SNRHa simply does not exceed our conservative threshold. In practice, therefore, all such stars either have weak or marginal Balmer emission in the XP spectra or are located in regions where the noise and continuum gradients make a robust detection difficult.
To illustrate this, in Fig.~\ref{fig:EmittersNotDetected} we plot the 
internal {\gaia} BP and RP spectra for 45 emitters that our algorithm did not detect (providing EWs equal to zero for all three Balmer lines). At the Balmer-line positions (vertical dashed lines), the emission lines reported in the literature are not visible in these spectra.

\section{Absorption line strength versus colour}
\label{sec:colourexcessOCs}

Equivalent widths (EWs) measure line strengths relative to the local continuum. Since hydrogen Balmer lines are narrow and the extinction curve is smooth, the EWs are essentially insensitive to interstellar extinction (see also \citealt{Dominguez2013} for EW-based analyses of Balmer lines). As a result, the EW versus colour relations for different clusters share the same intrinsic shape, but shifted in colour by different amounts depending on the cluster reddening (Fig.~\ref{fig:mastercurve}). This behaviour allows us to derive cluster colour excesses and absorption.

Our methodology uses all stars in a cluster to obtain a global estimate of the cluster excess without considering inner-cluster variations in extinction. Although significant differences occur within a cluster, they can easily be confused with other noise effects, and their detailed analysis is beyond the scope of this work.

\subsection{Empirical EW–colour relations (master curves)}

\begin{figure*}[!htbp]
    \centering
    \includegraphics[width=0.32\textwidth]{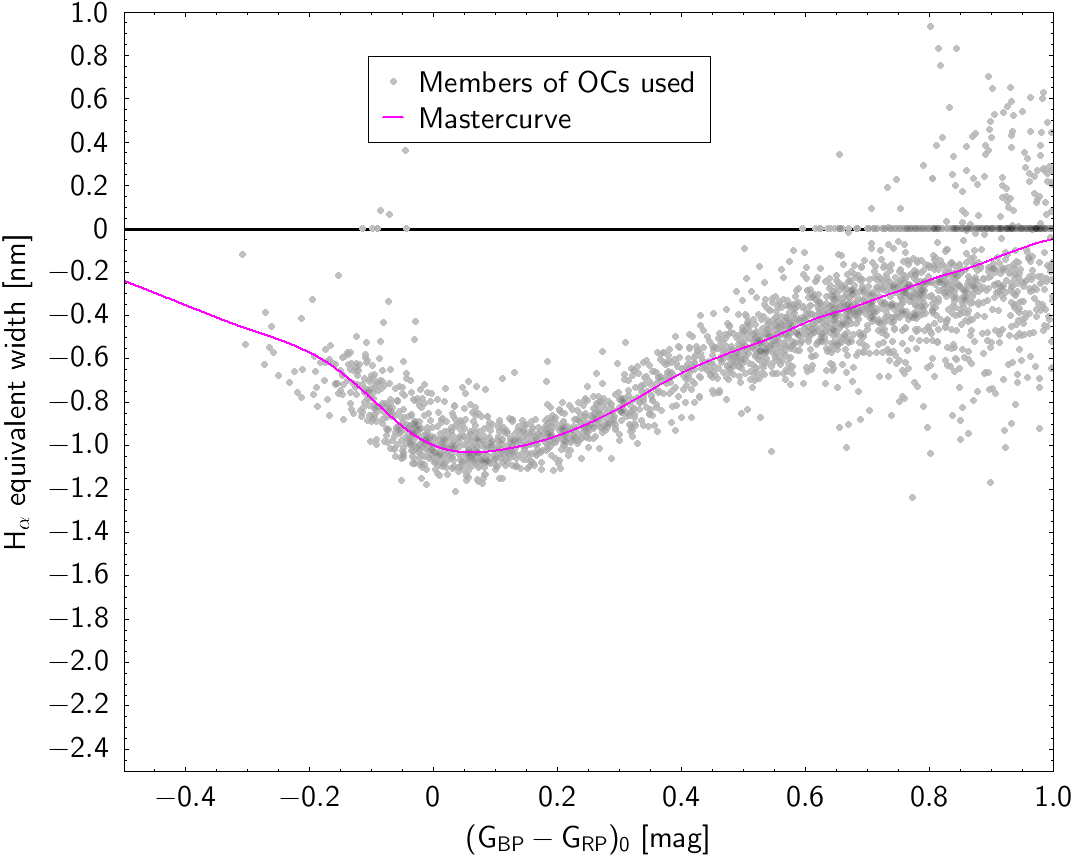}
    \includegraphics[width=0.32\textwidth]{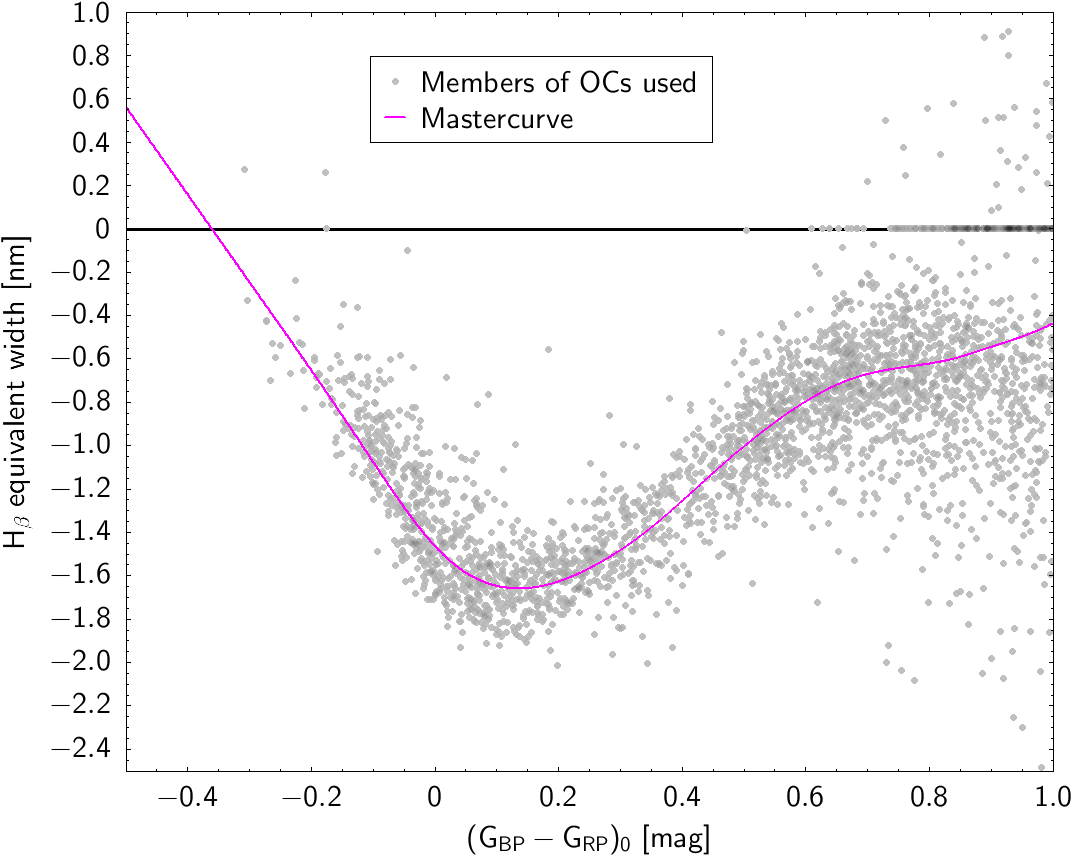}
    \includegraphics[width=0.32\textwidth]{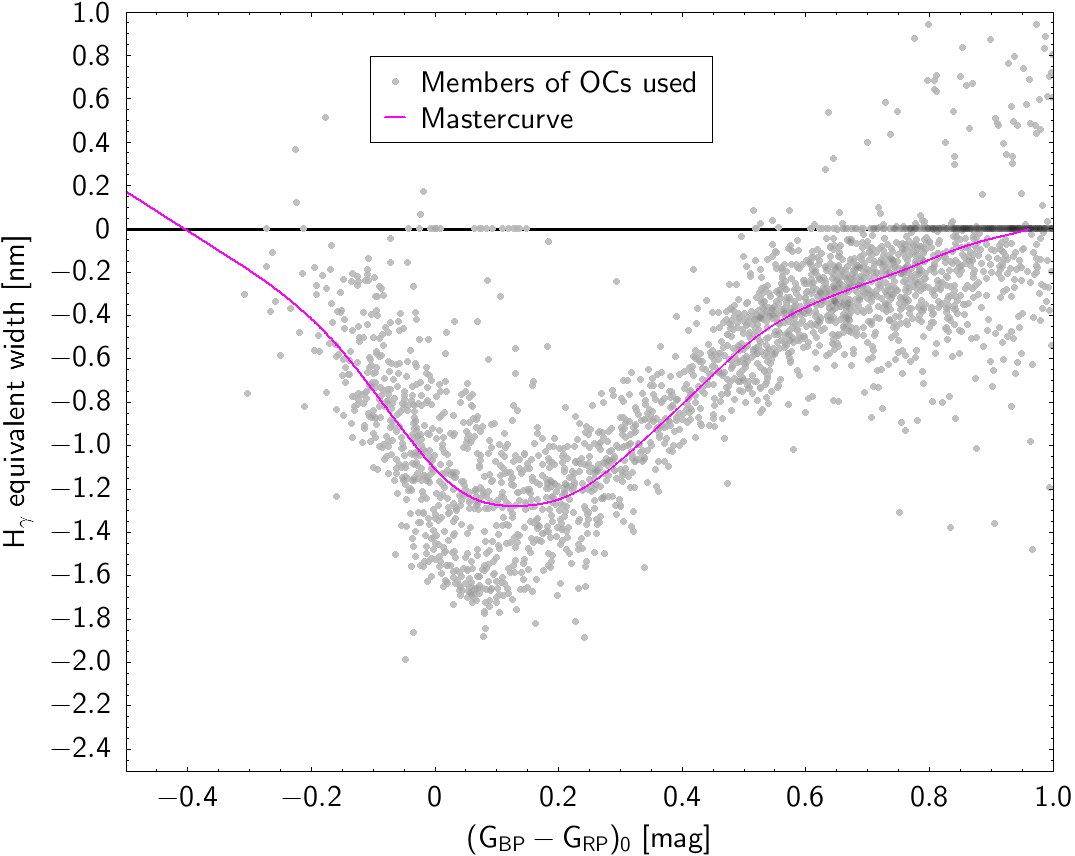}    
    \caption{
    Equivalent widths (EWs) of \Ha, \Hb, and \Hg\ versus intrinsic \bprp\ colour for members of nine OCs with negligible extinction. Grey points represent individual stars, and magenta lines show the fitted EW-colour master curves used to derive colour excesses and identify outliers with anomalous Balmer strengths.}
    \label{fig:mastercurve}
\end{figure*}

We constructed empirical EW–colour relations (using \bprp colour) for \Ha, \Hb, and \Hg\ for main-sequence stars selected from the CMDs of a set of reference OCs. The chosen OCs fulfil three criteria: (i) they have a large number of members; (ii) their Balmer lines have good S/N; and (iii) they are young enough to include hot blue stars. The nine clusters that best satisfy these criteria are NGC\,3532, NGC\,3114, NGC\,2287, NGC\,6475, NGC\,2451A, NGC\,2422, BH\,99, Trumpler\,10, and Pozzo\,1 (with 150-1541 members, mean S/N between 2 and 5, and minimum {\bprp} between -0.5 and -0.1~mag). For each OC we shifted the \bprp\ colours until the EW-colour relation matched that of the bluest reference cluster. We then fitted smoothing splines to the combined sample to obtain master curves for the three Balmer lines (Table~\ref{tab:mastercurve} and Fig.~\ref{fig:mastercurve}).

The master curves are reliable over \(-0.2 \lesssim (\bprp)_0 \lesssim 0.6\) mag. At bluer colours the number of stars is too small to constrain the relation, and at redder colours molecular bands complicate the interpretation of Balmer EWs. The curves reach minima (i.e. maximum absorption) at intrinsic colours of \((\bprp)_0 = 0.06\) mag for \Ha, \((\bprp)_0 = 0.136\) mag for \Hb, and \((\bprp)_0 = 0.13\) mag for \Hg, which correspond to effective temperatures of roughly 8500–9500\,K for main-sequence stars with \(\log g \approx 4.5\), i.e. spectral types close to A0V \citep{jordi2010}.

\begin{figure*}[!htbp]
     \centering
     \includegraphics[width=0.33\linewidth]{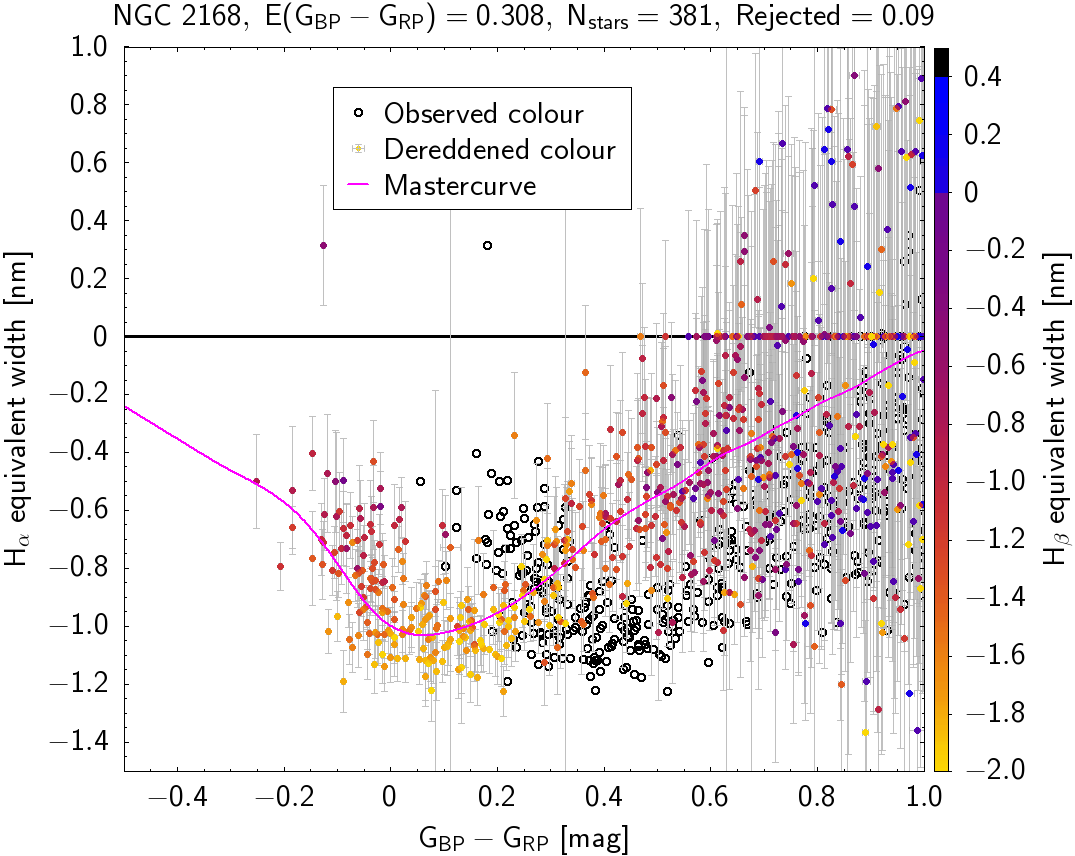}
     \includegraphics[width=0.33\linewidth]{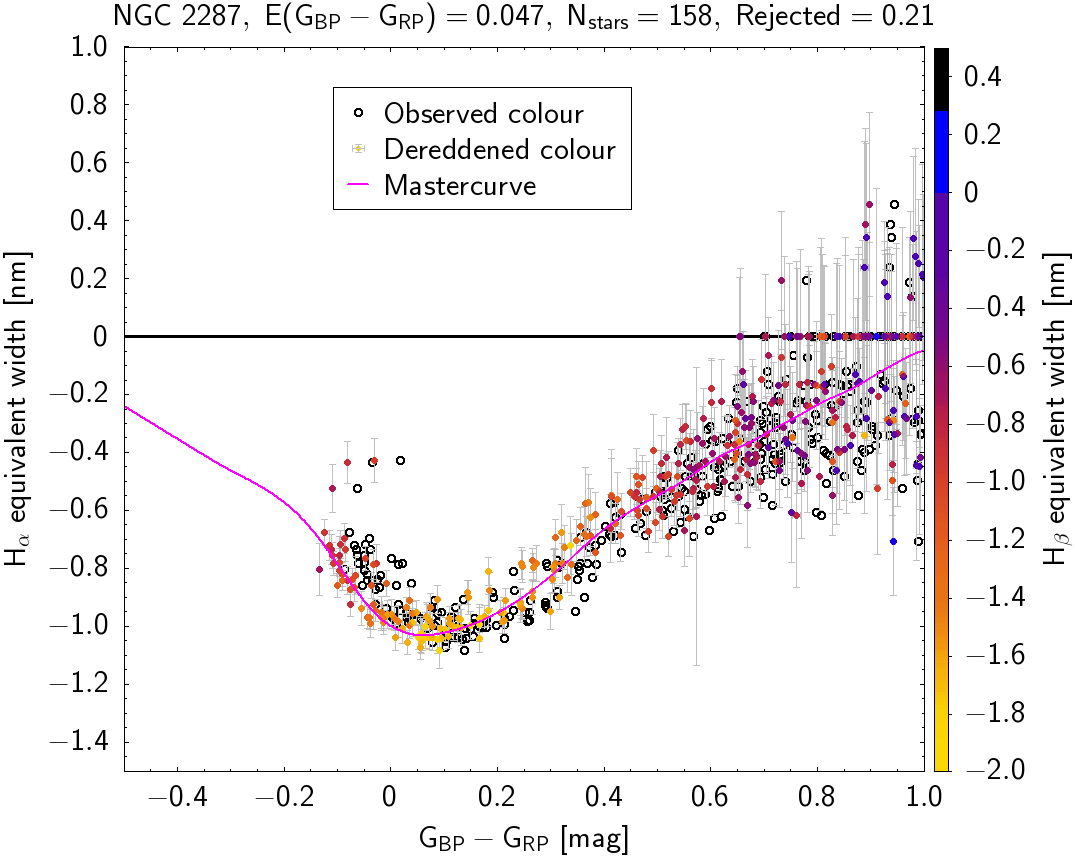}  
     \includegraphics[width=0.33\linewidth]{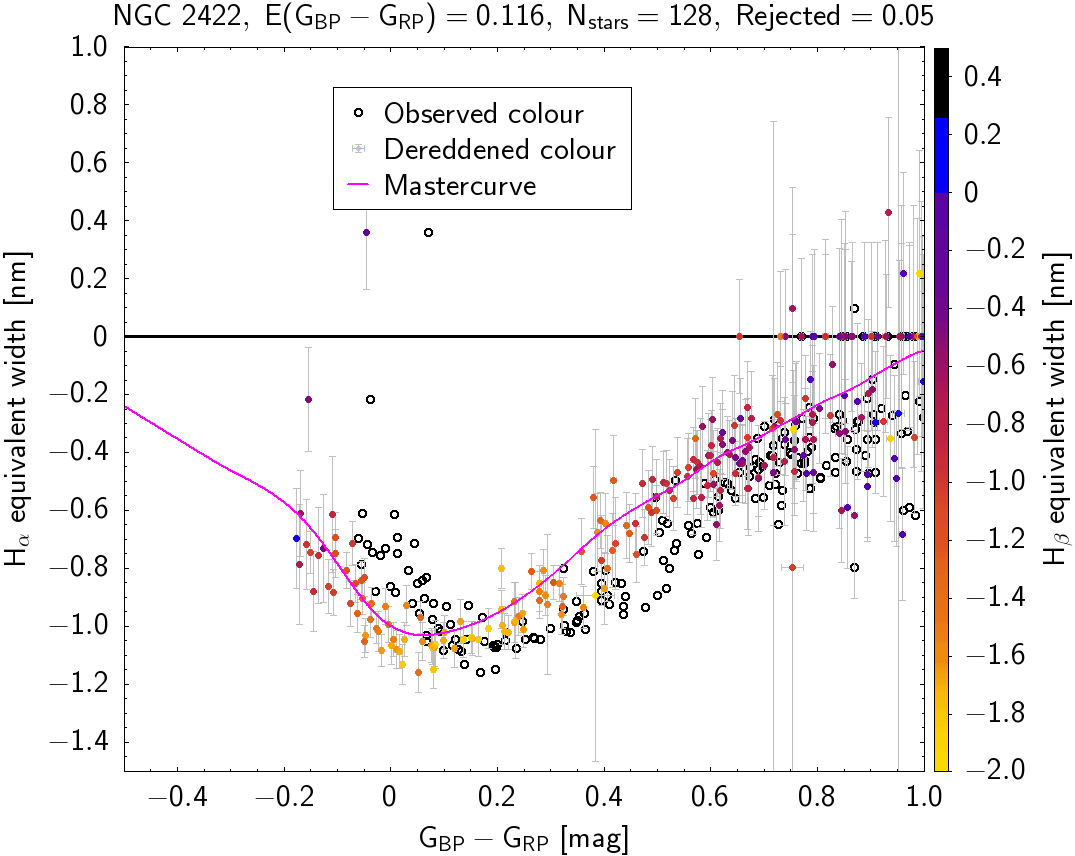}
     \includegraphics[width=0.33\linewidth]{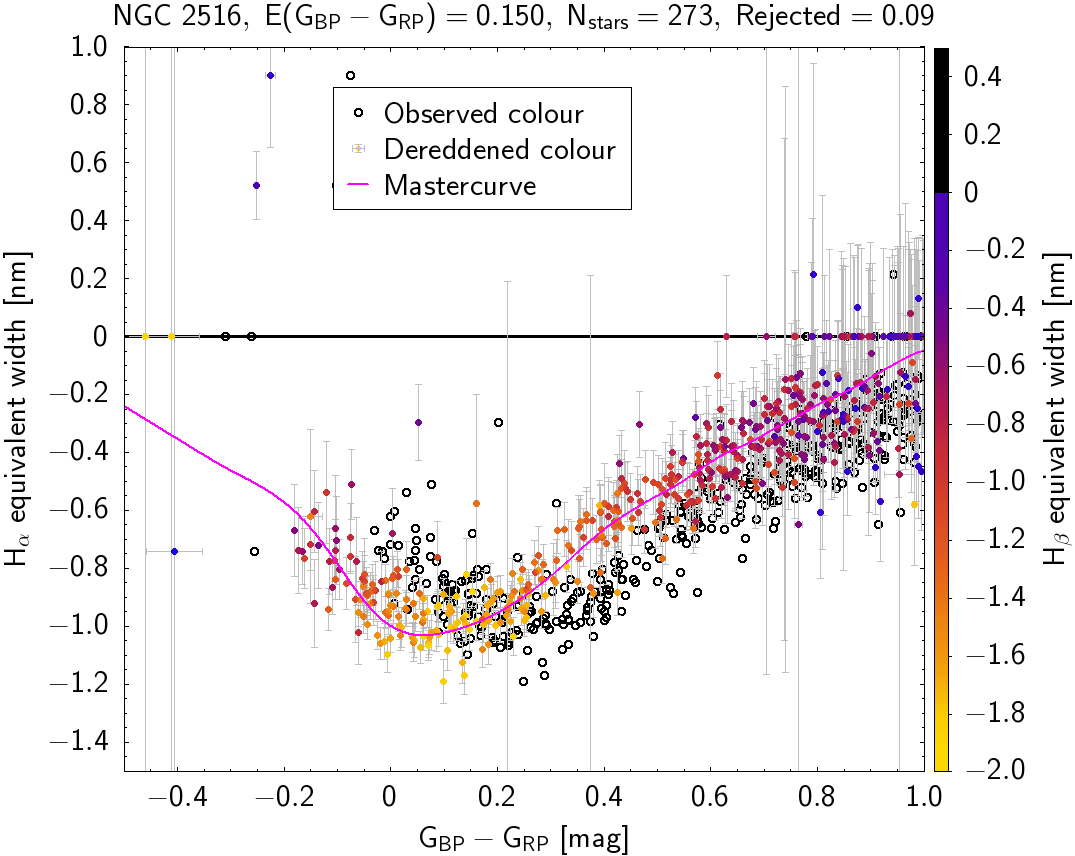}
     \includegraphics[width=0.33\linewidth]{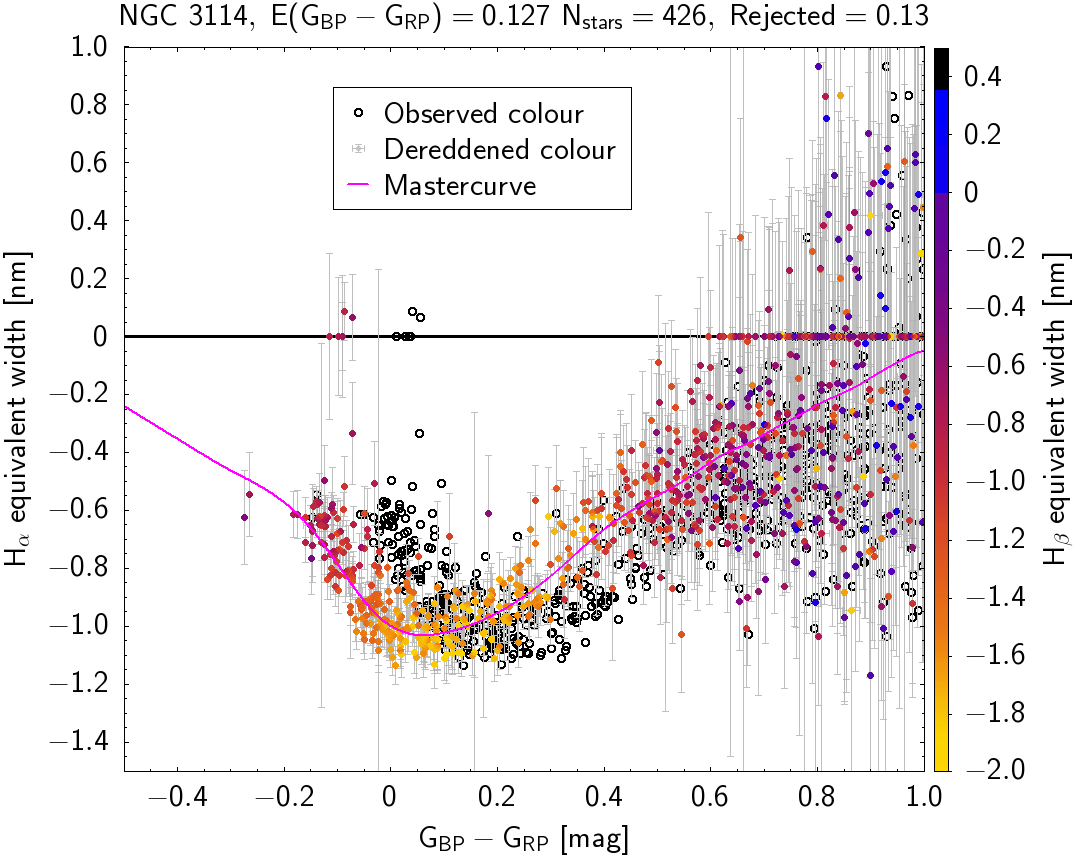}  
     \includegraphics[width=0.33\linewidth]{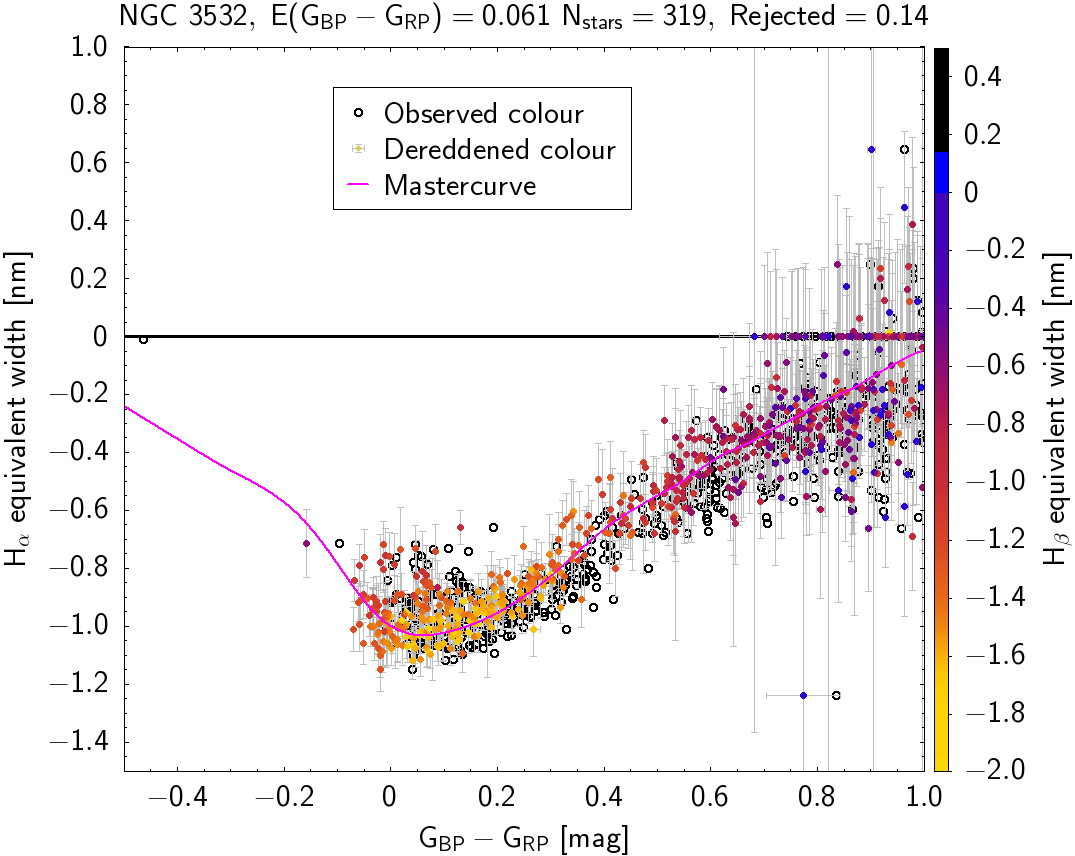}
    \caption{
     Relation between \Ha\ EW and \bprp\ for the same OCs as in Fig.~\ref{fig:cmd}. Open black symbols show stars plotted with their observed colours; filled symbols show the same stars after shifting by the colour excess obtained in Sect.~\ref{sec:colourexcessOCs}. Symbol colour encodes the \Hb\ EW. The title of each panel includes the values for the colour excess, the number of stars and the fraction (normalised to one) of outliers in the cluster.}
     \label{fig:HavsBPRPvsHb}
 \end{figure*}

Figure~\ref{fig:HavsBPRPvsHb} shows how Balmer EWs vary with colour for the same sample of OCs, highlighting the maximum absorption at \((\bprp)_0 \approx 0.1\) mag. We obtain the intrinsic colours used in this figure from the colour excesses derived below.

\subsection{Deriving OC colour excesses}
\label{sec:EBPRPdetermination}
We next used the master curves to derive colour excesses for all OCs. The excess colour derived here is only valid for stars with strong hydrogen absorption, that is, for stars of spectral types B and A. As the {\gaia} passbands are very wide, the excess colour for other spectral types, in particular for much cooler stars, is significantly different from the values derived here, even under the same conditions for extinction (see \citealt{jordi2010}).

For each OC we fitted the three master curves simultaneously to its EW–colour data by allowing a shift in \bprp. To mitigate the impact of outliers (notably emission-line stars; Sect. \ref{sec:emissionOCs}), we adopted an iterative procedure: we first determined the \bprp\ shift that minimises the error-weighted \(\chi^2\), then rejected data points deviating by more than three standard errors from the curves, and then repeated the procedure. The fits typically converge within a few iterations. We restricted the sample to stars that were bluer than the master-curve minimum minus 0.5 mag in \bprp\ to ensure that we worked in the regime where Balmer absorption was strong and the relations were well behaved. We took the final \bprp\ shift as the cluster colour excess \(E(\bprp)\). We estimated the uncertainty from the \(\chi^2\) curve as the half-width of the interval in colour shift for which \(\chi^2\) increases by 0.5 from its minimum \citep{NR}.

The master-curve for the \Ha line is plotted as a magenta line in Fig.~\ref{fig:HavsBPRPvsHb} (which is the same for all OCs). We included the values of the colour excess determined for the clusters in \cite{Hunt2023} in the online version of Table~\ref{tab:clusters}.
The grey data points in the CMD shown in Fig.~\ref{fig:cmd} represent the observed data and the coloured symbols show the same data shifted in colour using the colour excess derived here. 

The procedure described here provides negative excesses for some clusters, when their curves are located at lower colours than the reference master-curve or when the curve is not easily identified based on the emission lines behaviour. This occurs for 73 of the 7135 clusters in the sample. These clusters also show negative absorption values when derived from the colour excess in Sec.\ref{sec:AGestimates}.

\subsection{Absorption and excess estimations in other passbands}
\label{sec:AGestimates}

We verified the $E(\bprp)$ values derived in Sec.~\ref{sec:EBPRPdetermination} by comparing them with other estimates of interstellar extinction from other studies (see Sec.~\ref{sec:ExtinctionComparison}). To enable this comparison, we derived the corresponding absorptions in the \gaia\ \(G\) and Johnson \(V\) bands, \(A_G\) and \(A_V\), using BaSeL-3.1 synthetic spectra \citep{basel}, following the methodology of \cite{jordi2010}. We used the \gdr3\ \gaia\ passbands \citep{riello2021}, the Johnson \(V\) response from \citet{bessell2012}, and the extinction law of \citet{Cardelli1989}. We simulated stars with $0<A_{550}<11$~mag, where $A_{550}$ is the monochromatic extinction magnitude at wavelength $\lambda=550$~nm.

\begin{figure*}[!htbp]
    \centering
    \includegraphics[width=0.32\textwidth]{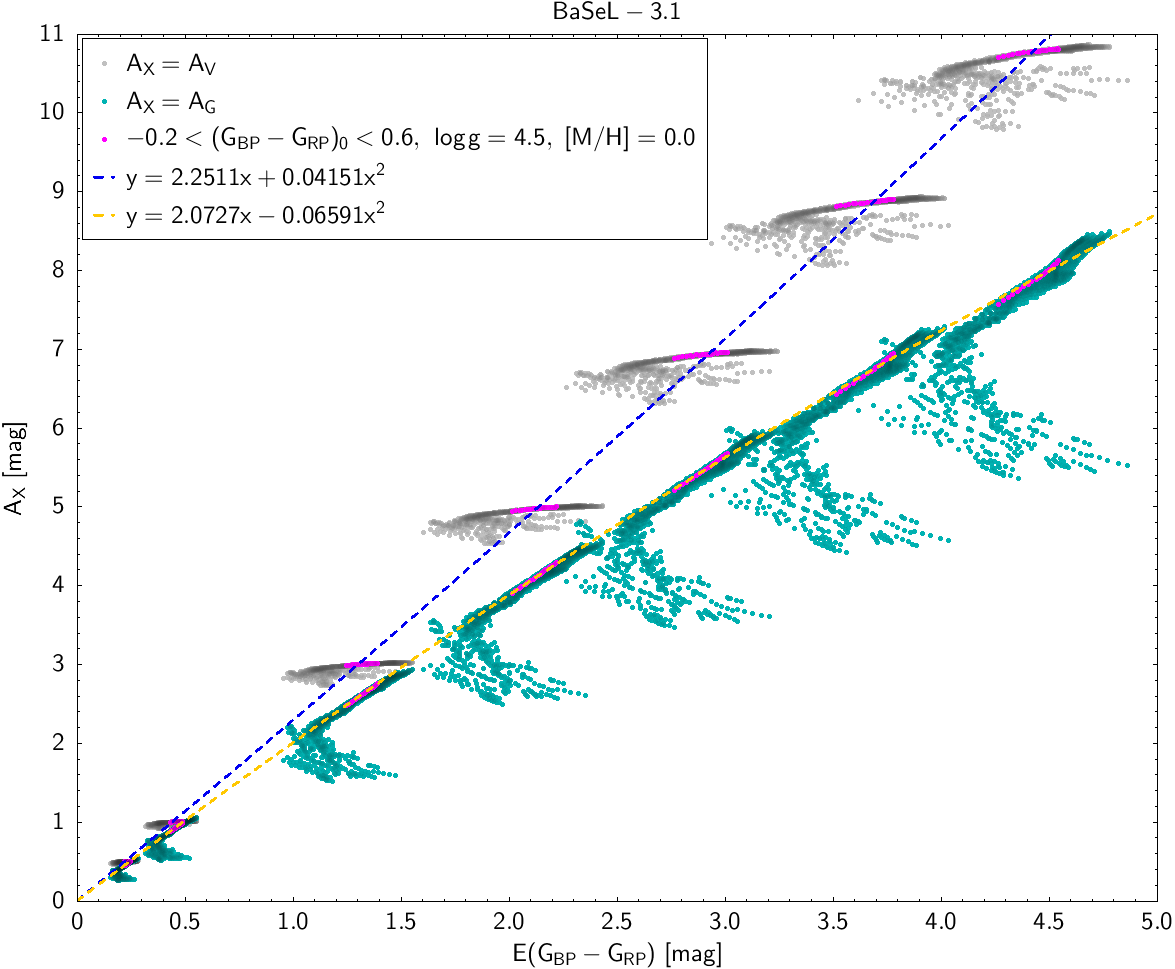}
    \includegraphics[width=0.32\textwidth]{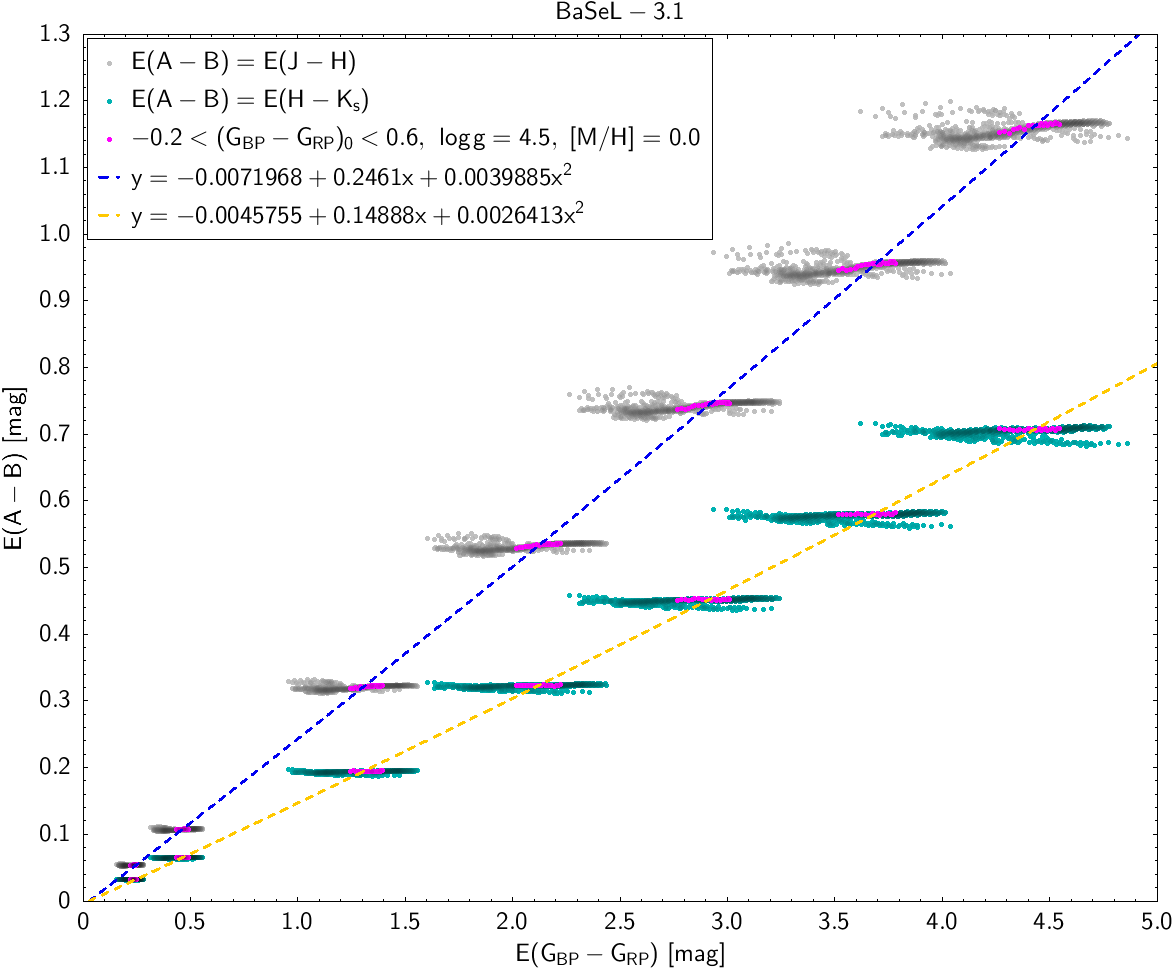}
    \includegraphics[width=0.32\textwidth]{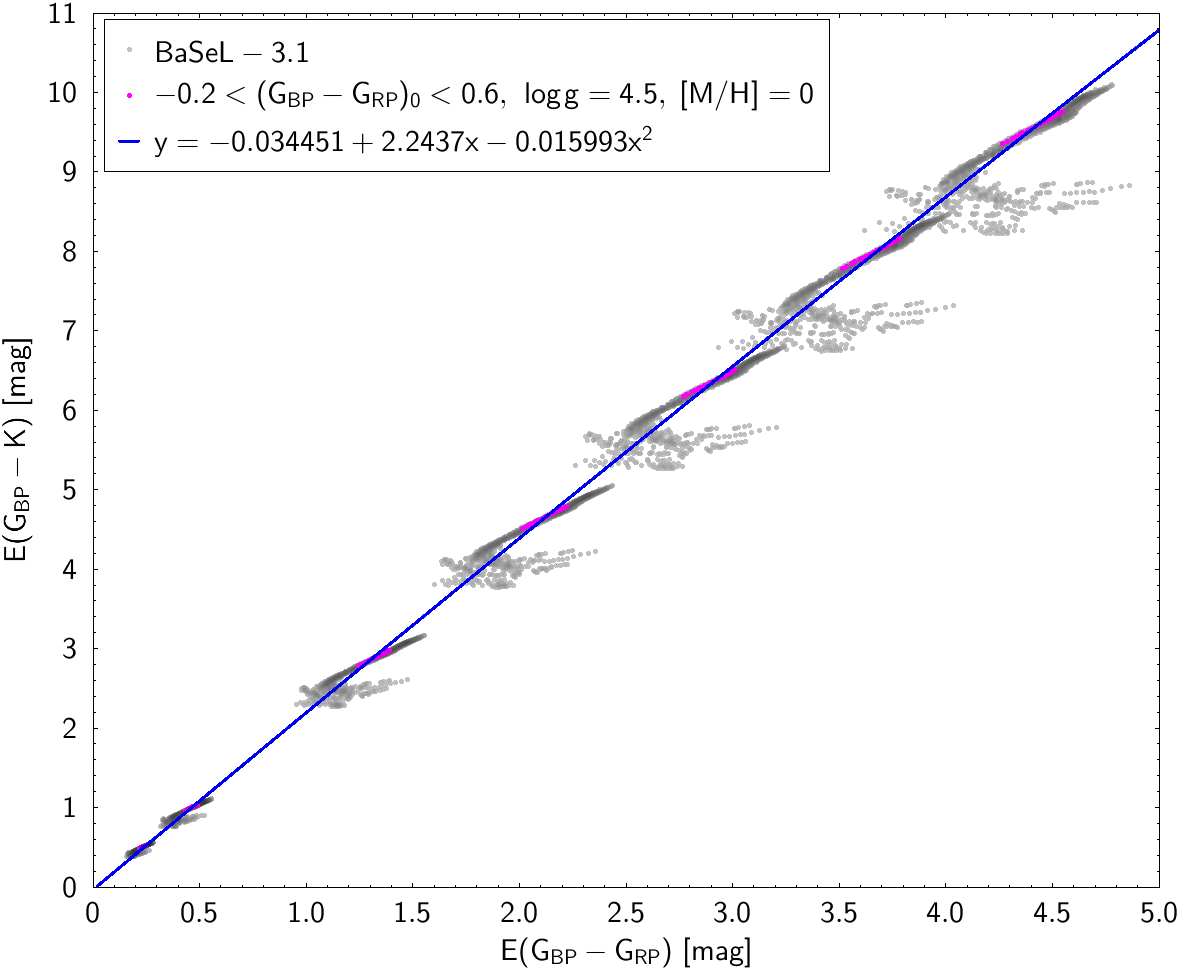}
    \caption{Simulations of absorption and colour excesses using the Basel-3.1 synthetic spectral library \citep{basel}. Left: Relationship between absorption in the $V$ (grey) and $G$ (cyan) passbands. For the polynomial fitting in Eqs.~\ref{eq:AVvsEBPRP_Basel} (blue line) and \ref{eq:AGvsEBPRP_Basel} (orange line), we used only the magenta points (sources with reliable $E(\bprp)$ estimations). Centre and right: Near-infrared 2MASS colour excesses as functions of \(E(\bprp)\) Centre: 2MASS \(E(J-H)\) (grey) and \(E(H-K_s)\) (cyan) and lines show the polynomial fits in Eqs.~\ref{eq:EJHvsEBPRP_Basel} and \ref{eq:EHKvsEBPRP_Basel}. Right: Excess in the combined optical–infrared colour \(E(G_{\rm BP}-K_S)\). The blue line shows the polynomial fit given in Eq.~\ref{eq:EBPKvsEBPRP}.    
    }
    \label{fig:AXvsBPRP}
\end{figure*}

We obtained empirical polynomial relations as a function of the colour excess $E(\bprp)$. We plot the derived polynomials in Fig.~\ref{fig:AXvsBPRP} (left) and list them in Eqs~\ref{eq:AVvsEBPRP_Basel} and \ref{eq:AGvsEBPRP_Basel} as follows:

\begin{equation}
\label{eq:AVvsEBPRP_Basel}
    A_V=2.2511E(\bprp)+0.04151E(\bprp)^2
\end{equation}

\begin{equation}
\label{eq:AGvsEBPRP_Basel}
    A_G=2.0727E(\bprp)-0.06591E(\bprp)^2
\end{equation}

We used the same simulations to relate our \gaia-based colour excesses to 2MASS colour excesses \citep{2MASSpassbands}. We adopted the following polynomial fits (Fig.~\ref{fig:AXvsBPRP}, centre and right panels):

\begin{eqnarray}
\label{eq:EJHvsEBPRP_Basel}
E(J-H)&=&-0.0071968+0.2461E(\bprp)+ \nonumber\\
      & &+0.0039885E(\bprp)^2
\end{eqnarray}

\begin{eqnarray}
\label{eq:EHKvsEBPRP_Basel}
    E(H-K_s)&=&-0.0045755+0.14888E(\bprp)+\nonumber\\
            & &+0.0026413E(\bprp)^2
\end{eqnarray}

To quantify the largest optical–infrared colour baseline we also derive a relation for \(E(G_{\rm BP}-K_s)\) (Fig.~\ref{fig:AXvsBPRP}, right panel),

\begin{eqnarray}  
\label{eq:EBPKvsEBPRP}
E(G_{BP}-K_S)&=&-0.034451+2.2437E(\bprp)+ \nonumber\\
             & &-0.015993E(\bprp)^2.
\end{eqnarray}

Using these equations, we derived the values for $A_V$, $A_G$, and the 2MASS excesses from our $E(\bprp)$ values in Sec.~\ref{sec:colourexcessOCs}. We include these estimates in Table~\ref{tab:clusters}.

\subsection{Comparison with other extinction estimates}
\label{sec:ExtinctionComparison}

This section compares colour excesses and absorption estimates derived in previous sections with estimates provided by other studies.

\begin{figure}[!htbp]
    \centering
    \includegraphics[width=0.8\columnwidth]{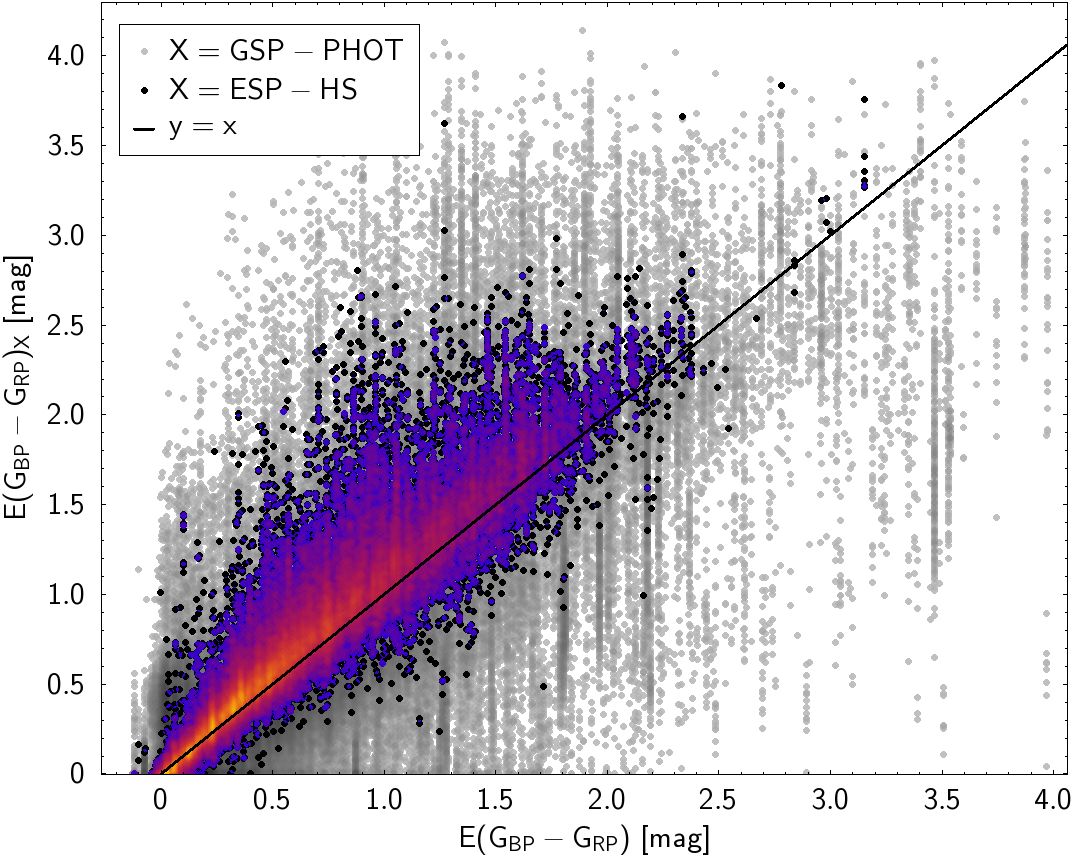}\\
    \caption{
    Colour excesses from the \gaia\ GSP-Phot \citep{GSPPHOT} and ESP-HS \citep{Creevey2022} pipelines (ordinate) versus the cluster \(E(\bprp)\) derived in this work (abscissa) for OCs. Only the ordinate changes between the two panels; the abscissa is always the Balmer-based \(E(\bprp)\) derived in Sect.~\ref{sec:colourexcessOCs}. The black line shows the identity relation.
    }
    \label{fig:EBPRPCU8vsEBPRP}
\end{figure}

First, we compared our cluster-averaged \(E(\bprp)\) with the star-by-star colour excesses provided by the \gaia\ GSP-Phot and ESP-HS pipelines \citep{GSPPHOT,Creevey2022}. Figure~\ref{fig:EBPRPCU8vsEBPRP} shows that the correlation is tighter with ESP-HS (which also uses the {\gaia} radial velocity spectrometer data, RVS) than with GSP-Phot (using only XP spectra), suggesting that our Balmer-based colour excesses are of comparable quality to those obtained from more complex models that also include RVS data.

\begin{figure}[!htbp]
    \centering
    \includegraphics[width=0.8\columnwidth]{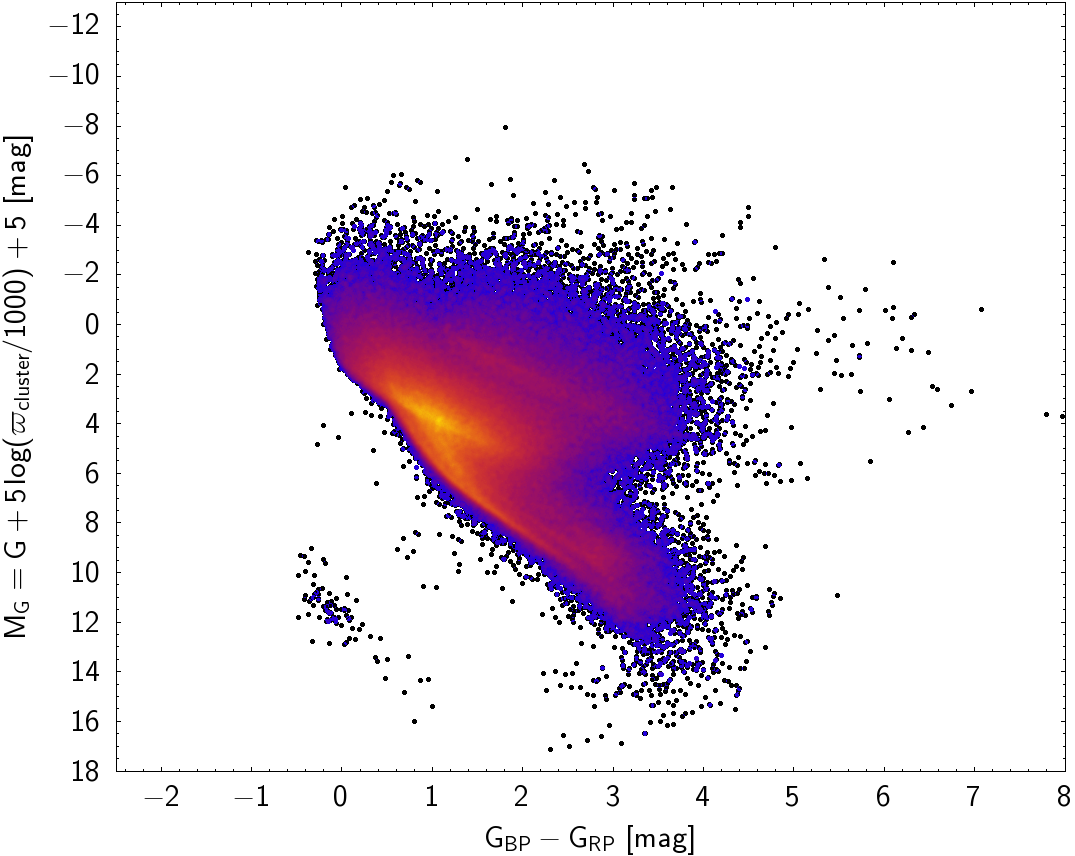}\\
    \includegraphics[width=0.8\columnwidth]{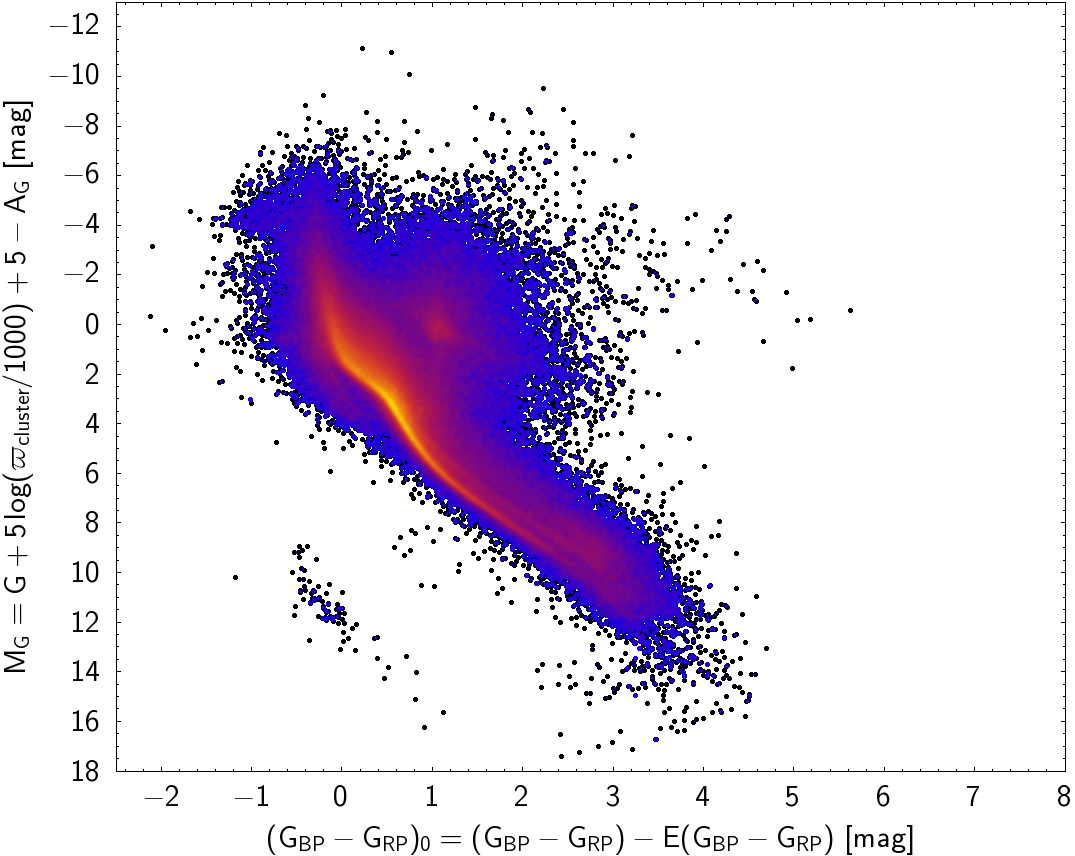}
    \caption{
    Colour-absolute magnitude diagrams (CAMDs) for all selected OC stars using observed \bprp\ colours (left) and intrinsic colours (right), obtained by subtracting the cluster $E(\bprp)$ and $A_G$ from BaSeL-3.1-based relations derived in Sect.~\ref{sec:colourexcessOCs}.
    }
    \label{fig:HRdiagram}
\end{figure}

Using {\gaia} colour excesses (Sec.~\ref{sec:colourexcessOCs}) and absorption in the {\gaia} $G$ passband (Eq.~\ref{eq:AGvsEBPRP_Basel} in Sec.~\ref{sec:AGestimates}), we de-reddened the CMD. Figure~\ref{fig:HRdiagram} illustrates the impact of our de-reddening procedure on the global CMD built from all OC members. After correcting the colours using our \(E(\bprp)\) estimates (Table~\ref{tab:members}) and applying the corresponding \(A_G\), the main sequence becomes narrower, the giant and binary sequences are better defined, and the morphology of the diagram is significantly cleaner.

\begin{figure}[!htbp]
\centering
    \hspace{-0.5
    cm}\includegraphics[width=0.61\columnwidth]{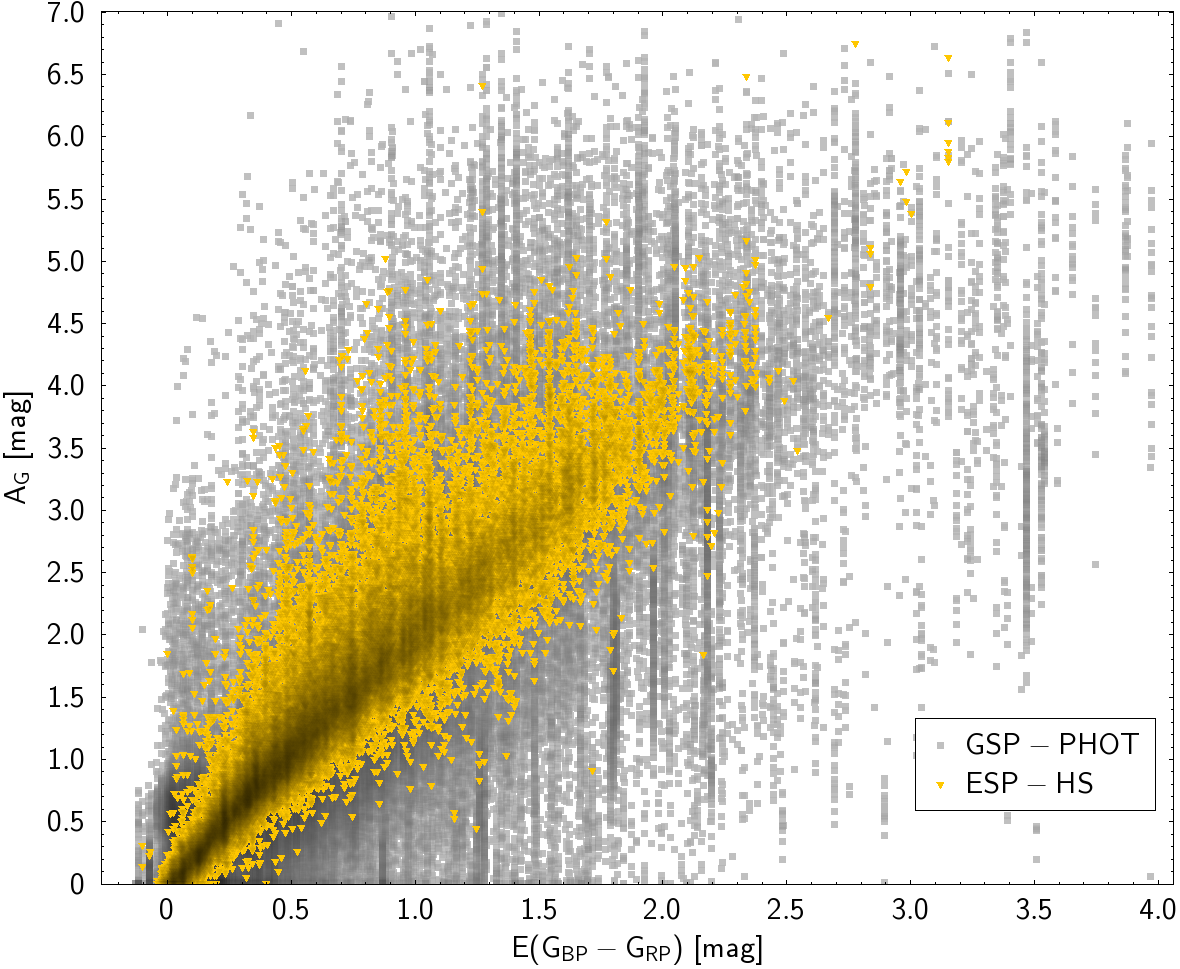}\\
    \includegraphics[width=0.7\columnwidth]{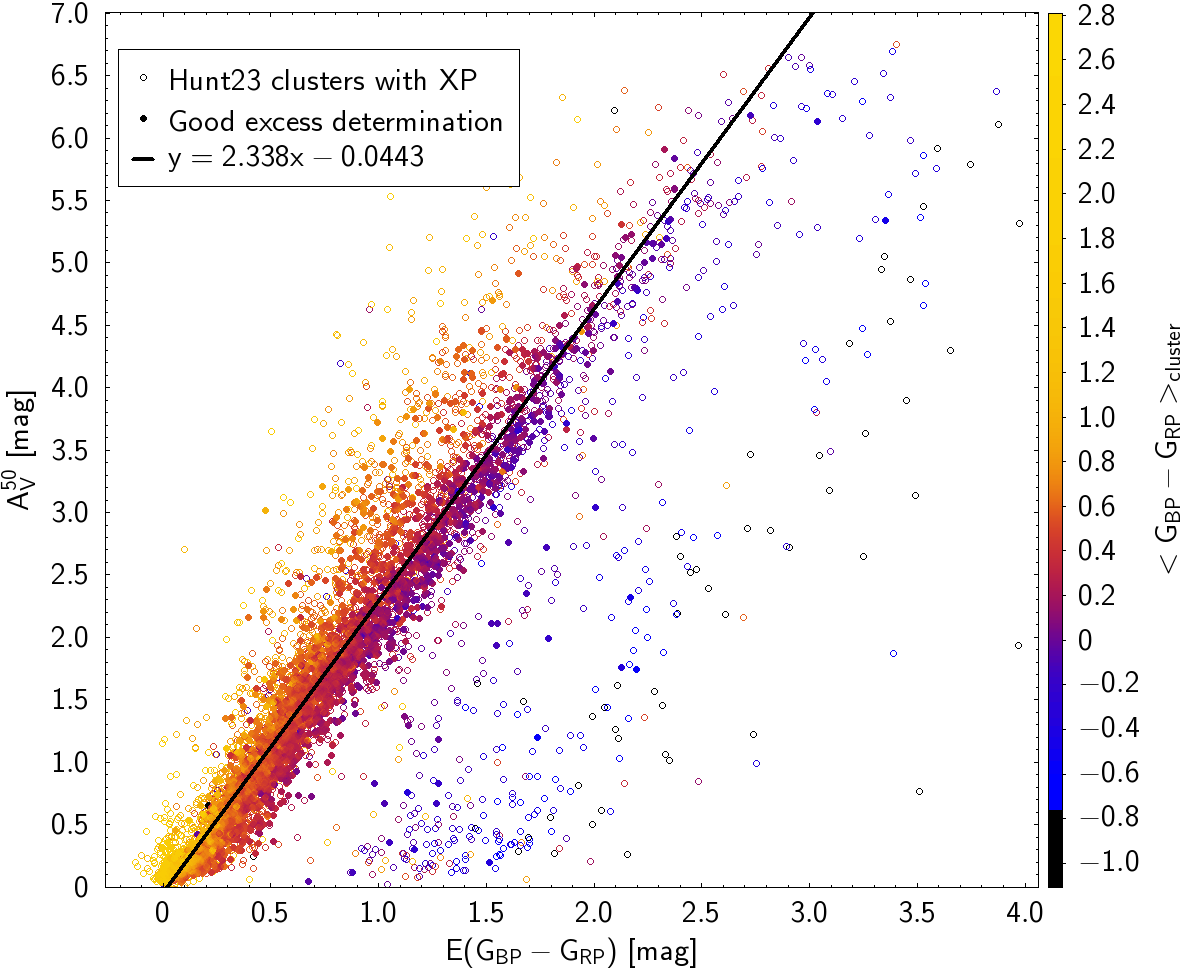}
    \caption{Interstellar absorption from different works plotted against the $E(\bprp)$ colour excess derived in this work for OCs (see Sect.~\ref{sec:colourexcessOCs}). 
    Top: Correlation between our colour excesses compared with the interstellar absorption derived from {\Gaia} GSP-Phot \cite{GSPPHOT} (grey squares) and ESP-HS \citep{Creevey2022} (orange triangles). Bottom: $A^{50}_V$ values in the ordinate axis from \cite{Hunt2023}, with solid symbol marking sources with good excess determination. The colour index indicates the median intrinsic colour of each OC. The black line shows a linear fit. Both panels use the same axis scales.}
    \label{fig:AVvsEBPRP}
\end{figure}

We expect that the colour excess is linearly related to the interstellar absorption (see Fig.~\ref{fig:AXvsBPRP} for the behaviour of the spectral library), with the slope given by the interstellar reddening\footnote{Nevertheless, some variation with source colour is also present (see Fig.~\ref{fig:AXvsBPRP}).}. 
This almost linear relationship is expected even when mixing passbands from different photometric systems\footnote{The advantage of comparing colour excesses and absorptions from different photometric systems is to minimise the chain of transformations required to compare them, thereby reducing systematic effects.}, but with a different slope (see left panel in Fig.~\ref{fig:AXvsBPRP}). For example, in Fig.~\ref{fig:AVvsEBPRP} (bottom) we compare our $E(\bprp)$ values with the interstellar absorption in the Johnson-Cousins $V$ band ($A_V^{50}$, the 50th percentile of V-band cluster extinction obtained using a Bayesian neural network) from \cite{Hunt2023} for the same cluster, showing that our estimated colour excesses are clearly correlated with $A_V^{50}$. 

We find a weaker correlation (see Fig.~\ref{fig:AXvsBPRP}, top) when using the individual $A_G$ 
values derived by the {\Gaia} collaboration \citep{GSPPHOT} for each member of the OC\footnote{We also used $A_0$ GSP-SPEC and MSC values to check whether the correlation improved, but they showed similar results to those obtained with $A_G$ shown in Fig.~\ref{fig:AVvsEBPRP} (top).}.

The better agreement of our colour excesses with \cite{Hunt2023} than with \cite{GSPPHOT} (see Fig.~\ref{fig:AVvsEBPRP}) is probably due to the fact that both \cite{Hunt2023} and our work use combined information from all sources in the OC, whereas \cite{GSPPHOT} performs a star-by-star analysis without using additional information from other stars in the cluster. We obtain a better correlation (although still worse than for the \citealt{Hunt2023} values) using the values from the {\gaia} Extended Stellar Parametrizer for Hot Stars (ESP-HS, \citealt{Creevey2022}, orange points in Fig.~\ref{fig:AVvsEBPRP}, top) which, in addition to XP information, uses data from the RVS. 

Most of the stars in the top panel of Fig.~\ref{fig:AXvsBPRP} are located close to the linear fit, although there are some outliers with larger $E(\bprp)$ values. This secondary sequence corresponds to inaccurate excess flux estimates caused by poorly shaped for the observed curves of hydrogen EWs as a function of colour or the low number of sources used for the fitting. To clean this diagram, we applied the following filters:

\begin{itemize}
    \item To ensure agreement between the original fit of the master curve and the minimum chi-squared as a function of the shift in colour, we only used clusters with disagreement smaller than $|E(\bprp)_{\rm corr}-E(\bprp)|<2\cdot 10^{-5}$~mag. The small threshold value reflects the fact that the two fitting methods used to derive the colour excess provide essentially the same result.
    \item We also ensured that a significant number of stars in each cluster was available for the fit by imposing $N_{\rm stars}>10$.
    \item We excluded clusters with too many outliers so that the fit remained meaningful. We enforced this by selecting only clusters with fewer than 15\% outliers (rejectedFraction$<0.15$).
    \item The last criterion ensured that the confidence interval was narrow ($\Delta(E(\bprp))<0.1$~mag), i.e. the value of the colour excess was well constrained. We estimated the uncertainty of the colour excess by shifting the master curves over the observed EW-\bprp relation on a fixed grid of shifts and computing the chi-squared value for each shift. We then used an interpolating spline to estimate the continuous chi-squared function with shift. We defined the upper and lower limit of the confidence interval as the shifts for which the chi-squared value increased with respect to the minimum value by a fixed amount. We adopted a value of 0.5, which corresponds to the 1$\sigma$ uncertainty interval for normally distributed errors.
\end{itemize}

This procedure yields \(E(\bprp)\) values for 6813 clusters of the 7135 with XP spectra analysed in Sect.~\ref{sec:clusters}.
 After applying additional quality cuts (e.g. on the fraction of rejected points and on the width of the confidence interval), we flag 2723 clusters as having reliable colour excesses ('GoodExcess' in Table~\ref{tab:clusters}). 
 Nevertheless, despite the filtering, a few OCs still lie in the secondary sequence. To provide the fitted line shown in Fig.~\ref{fig:AVvsEBPRP} (bottom), we added an extra filter, using only sources with $A_V>2.4E(\bprp)-1$. This mild filter leaves 2671 OCs, which we used to derive a global relation between \(E(\bprp)\) and the median \(V\)-band absorption from \citet{Hunt2023}:

\begin{equation}
\label{eq:AVvsEBPRP_Hunt}
A_V^{50}=2.338 E(\bprp)-0.0443. 
\end{equation}

This relation captures the mean trend for all clusters, but the slope depends to some extent on the median intrinsic cluster colour (redder clusters exhibit larger slopes, bluer clusters smaller ones; see Fig.~\ref{fig:AVvsEBPRP}, bottom).

Using the OCs in common between \citet{Hunt2023} and \citet{Cantat2020Portrait}, we also compared their \(A_V\) values as functions of our \(E(\bprp)\). When we adopt \(A_V\) from \citet{Cantat2020Portrait}, we obtain a shallower relation:
\begin{equation}
\label{eq:AVvsEBPRP_Cantat}
A_V^{50}=1.8684E(\bprp)+0.0409
\end{equation}
The differences between the absorption scales of \citet{Hunt2023} and \citet{Cantat2020Portrait} are discussed in detail by \citet{Hunt2023}, 
who attribute them mainly to differential extinction of stars inside the clusters.

We find that the slope recovered from our simulations (2.2511, Eq.~\ref{eq:AVvsEBPRP_Basel}) for the $A_V=f(E(\bprp))$ relation is closer to that obtained using the \cite{Hunt2023} data (2.338, Eq.~\ref{eq:AVvsEBPRP_Hunt}) than that derived from \cite{Cantat2020Portrait} (1.8684, Eq.~\ref{eq:AVvsEBPRP_Cantat}).

\begin{figure}[!htbp]
    \centering
    \includegraphics[width=0.8\columnwidth]{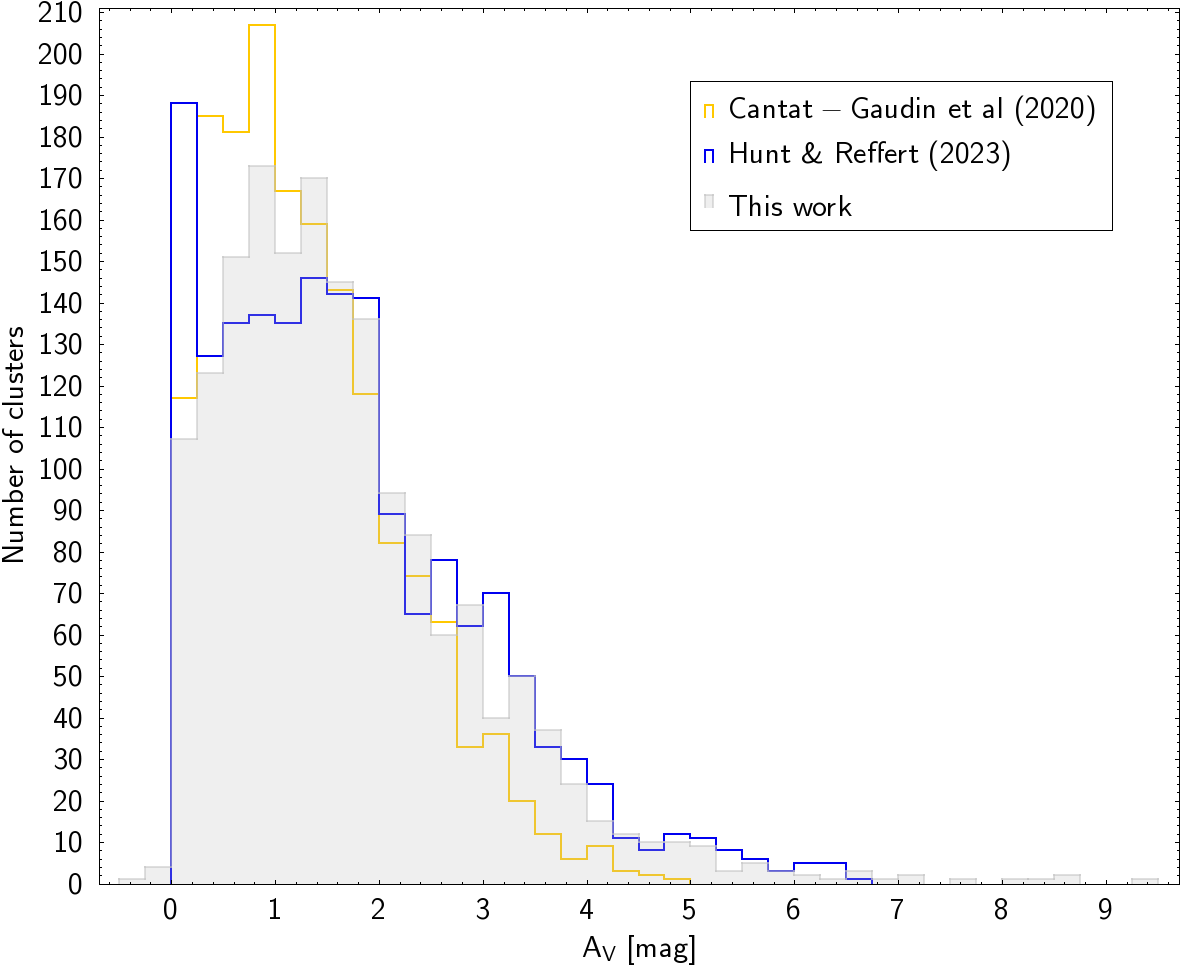}
    \caption{Histogram of $A_V$ values included in \cite{Cantat2020Portrait} (orange), \cite{Hunt2023} (blue), and derived using Eq.~\ref{eq:AGvsEBPRP_Basel} for clusters in common with $A_G>0$~mag from Eq.~\ref{eq:AGvsEBPRP_Basel}. 
    }
    \label{fig:HistogramAV_HuntCantatBasel}
\end{figure}

Figure~\ref{fig:HistogramAV_HuntCantatBasel} compares the distribution for the clusters in common between \cite{Hunt2023} and \cite{Cantat2020Portrait}, considering only positive values for $A_G$ as derived by Eq.~\ref{eq:AGvsEBPRP_Basel}. Our $A_V$ values are more similar to those of \cite{Hunt2023}, although we obtain more sources with $A_V<2$~mag.

\begin{figure}[!h]
\centering
    \includegraphics[width=0.8\columnwidth]{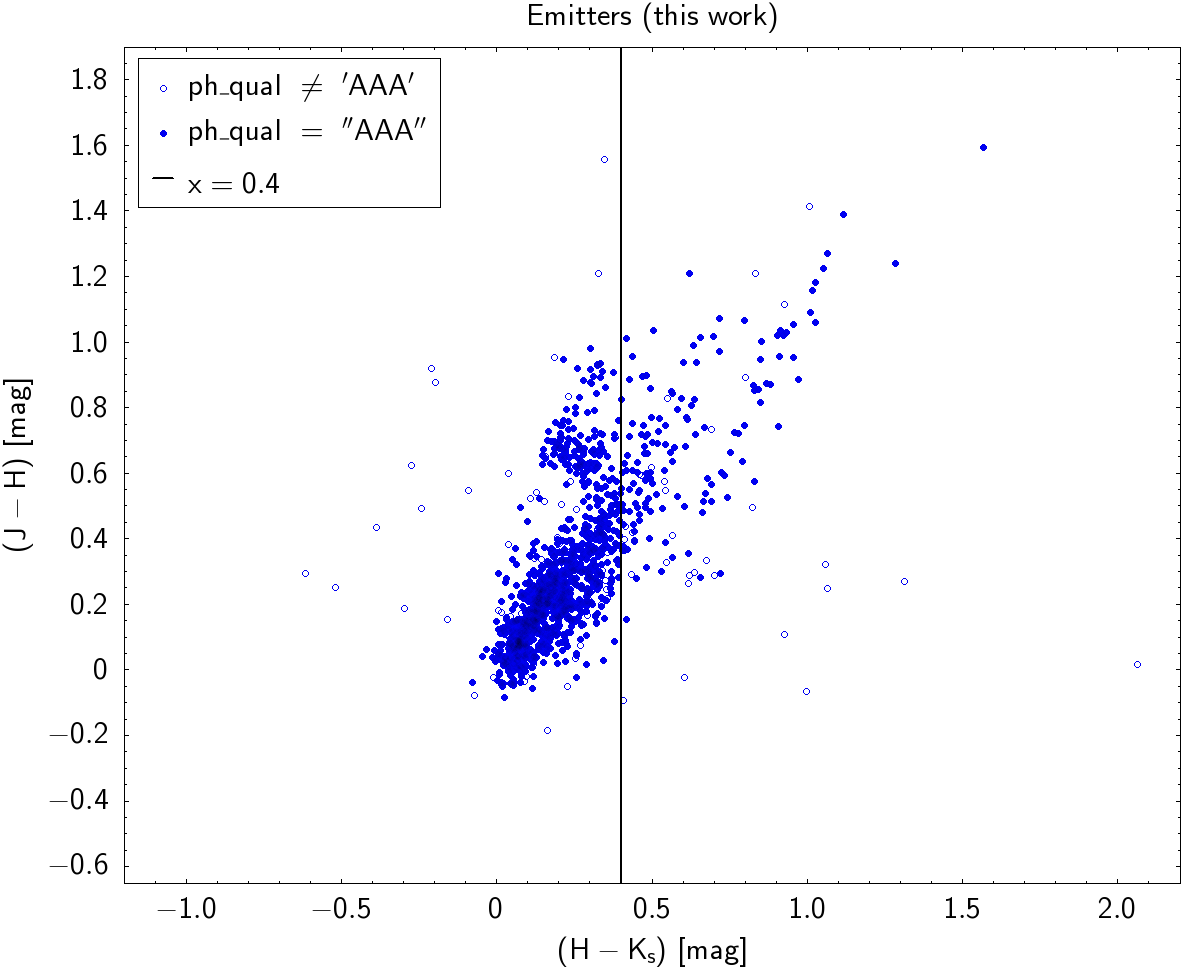}
    \includegraphics[width=0.8\columnwidth]{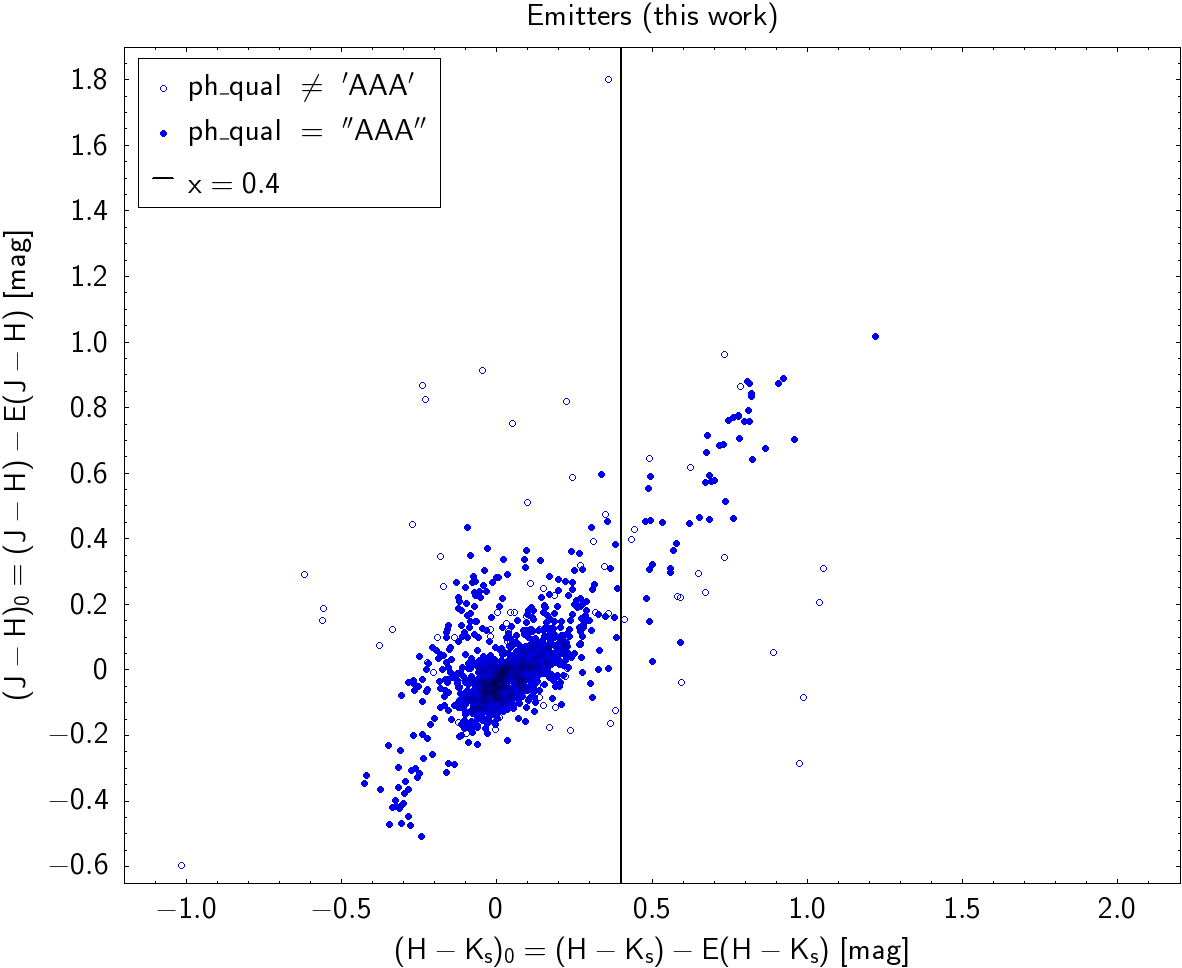}
    \caption{
    Observed (top) and intrinsic (bottom) 2MASS colour–colour diagrams for the emission-line stars identified in this work. Filled symbols mark sources with 2MASS photometric quality `AAA' in all bands. Intrinsic colours are derived using Eqs.~\ref{eq:EJHvsEBPRP_Basel} and \ref{eq:EHKvsEBPRP_Basel}.}
    \label{fig:JHvsHK}
\end{figure}

We also examined the near-infrared colours of our emission candidates using 2MASS photometry \citep{2MASSpassbands} and the \gdr3–2MASS crossmatch \citep{Marrese2019}. \cite{Nidhi2023}, using 2MASS colours \citep{2MASS}, adopts the criterion $(H-K_s)_0>0.4$~mag to select Herbig Ae/Be stars from classical Be and Ae stars. Of the 1256 emission candidates studied in Sect. \ref{sec:emissionOCs}, 1228 have 2MASS counterparts, and 1108 of these have photometric quality flag `AAA' (with $S/N>10$ in all three 2MASS passbands). Using Eqs.~\ref{eq:EJHvsEBPRP_Basel} and \ref{eq:EHKvsEBPRP_Basel} to remove reddening effects, we constructed intrinsic 2MASS colour–colour diagrams (Fig.~\ref{fig:JHvsHK}). We identify 70 stars (51 with `AAA' quality) with intrinsic \((H-K_s)_0 > 0.4\) mag, making them good candidates for Herbig Ae/Be stars.

\section{Summary and conclusions}
\label{sec:conclusions}

We analysed the \gdr3\ internal XP spectra of stars in stellar clusters from \citet{Hunt2023} using the Hermite-function formalism of \citet{weiler2023} to measure EWs for the Balmer lines \Ha, \Hb, and \Hg. Open clusters (OCs), whose members share common ages, metallicities, distances, and reddening, provide an ideal test-bed for assessing the sensitivity and robustness of our method. The empirical Balmer EW–colour relations derived from a set of reference OCs define master curves that are nearly insensitive to extinction. We used these to obtain cluster colour excesses and to identify stars with anomalous Balmer strengths.

From these measurements we obtain \(E(\bprp)\) colour excesses for more than 6800 clusters, derive empirical relations linking \(E(\bprp)\) to \(A_G\), \(A_V\), and 2MASS colour excesses, and demonstrate that the resulting de-reddened CMDs show significantly tightened main sequences and clearer cluster features. We also identify 1256 emission-line candidates (1157 in OCs), of which 356 (312 in OCs) appear to be new compared to ESP-ELS and SIMBAD. Finally, the XP-based colour excesses show good agreement with independent estimates from ESP-HS, confirming that Balmer-line strengths in \gaia\ XP spectra can provide reliable reddening indicators when applied to suitable samples of stars.

We have demonstrated that despite the very low resolution of the XP spectra, useful line information can be extracted, and the large number of sources can yield high quality scientific results. 

Our next steps are to extend this approach to the full \gdr3\ catalogue and to additional spectral features, such as the Ca\,II triplet in white dwarfs, helium lines in hot stars, and molecular bands in cool dwarfs and carbon stars.

\section*{Data availability}

Tables~\ref{tab:clusters}--\ref{tab:mastercurve} are available in electronic form at the CDS via anonymous ftp to cdsarc.u-strasbg.fr (130.79.128.5) or via http://cdsweb.u-strasbg.fr/cgi-bin/qcat?J/A+A/

\begin{acknowledgements}
This work was (partially) supported by the Spanish MICIN/AEI/10.13039/501100011033 and by ''ERDF A way of making Europe'' by the “European Union” through grant PID2021-122842OB-C21 and PID2024-157964OB-C21, and the Institute of Cosmos Sciences University of Barcelona (ICCUB, Unidad de Excelencia ’Mar\'{\i}a de Maeztu’) through grant CEX2024-001451-M and the project 2021-SGR-00679 GRC de l'Agència de Gestió d'Ajuts Universitaris i de Recerca (Generalitat de Catalunya). 

This work was partially supported by the OCRE awarded project Galactic Research in Cloud Services (Galactic RainCloudS). OCRE receives funding from the European Union’s Horizon 2020 research and innovation programme under grant agreement no. 824079.\\

This work has made use of data from the European Space Agency (ESA) mission {\it Gaia}\footnote{\href{https://www.cosmos.esa.int/gaia}{https://www.cosmos.esa.int/gaia}}, processed by the {\it Gaia} Data Processing and Analysis Consortium (DPAC)\footnote{\href{https://www.cosmos.esa.int/web/gaia/dpac/consortium}{https://www.cosmos.esa.int/web/gaia/dpac/consortium}}. Funding for the DPAC has been provided by national institutions, in particular the institutions participating in the {\it Gaia} Multilateral Agreement.

This research has made use of the SIMBAD\footnote{http://simbad.cds.unistra.fr/simbad/0} database, operated at CDS, Strasbourg, France.

This research has made use of the TOPCAT\footnote{\href{http://www.starlink.ac.uk/topcat/}{http://www.starlink.ac.uk/topcat/}} and STILTS tools \citep{topcat}.

We also thank R. Sordo for her insightful comments on the manuscript, which helped improve the paper.

\end{acknowledgements}

\bibliographystyle{aa} 
\bibliography{SpectralLines}

\begin{appendix}
\nolinenumbers

\section{XP-TEAL}
 \label{sec:xpteal}

\begin{figure*}[!htbp]
    \centering
    \includegraphics[width=0.99\linewidth]{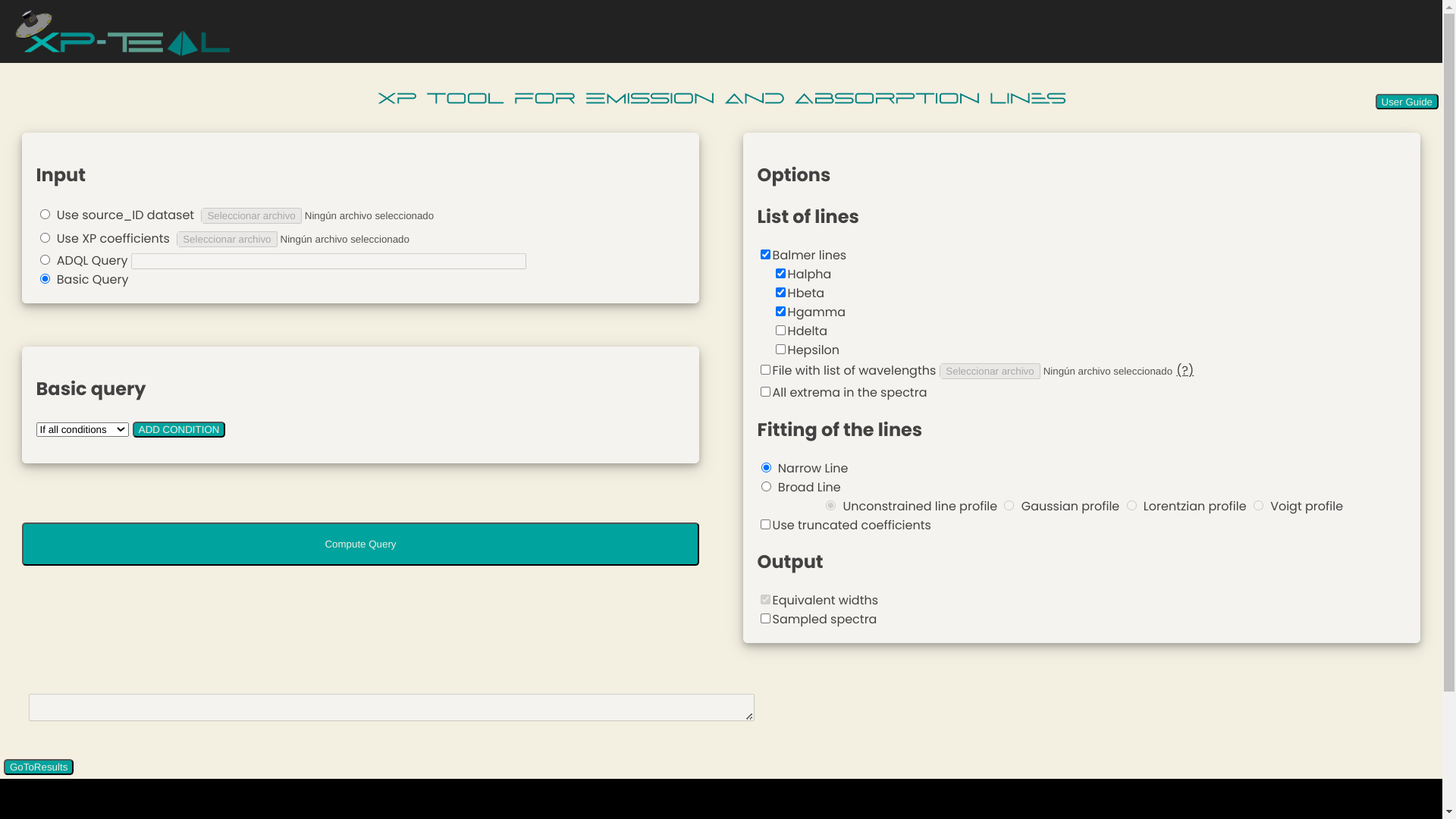}
    \caption{Screenshot of XP-TEAL web tool
    that could be used by the reader to run the same code used in this work.}
\footnotetext{\url{https://xpteal.fqa.ub.edu}}
    \label{fig:xpteal}
\end{figure*}

XP-TEAL can be used through a web interface\footnote{\url{https://xpteal.fqa.ub.edu}} (see screenshot in Fig.\ref{fig:xpteal}). The underlying Python source code is also publicly available\footnote{\url{https://github.com/malhotrasagar15/xpy_teal}}. It implements the formalism of \citet{weiler2023} for the analysis of emission and absorption lines in internally calibrated \gdr3\ XP spectra, exploiting the Hermite-function representation to locate spectral features efficiently and to avoid deconvolution artefacts.

As explained in Sect.~\ref{sec:method}, the method searches for local extrema in the internally calibrated spectra or in their second derivatives, and provides several options for computing EWs. When no significant extremum compatible with a spectral line is found, XP-TEAL returns upper or lower limits for the EW. Users should, however, carefully inspect the results: not every extremum is a genuine spectral line, even if its position in wavelength matches with the expected position of a line. Their interpretation must always take into account the expected spectral properties of the targets and the noise characteristics of the data.
The web interface of the tool is directly using {\gdr3} XP spectra from the {\gaia} archive\footnote{\url{https://gea.esac.esa.int/archive/}}. Four different options for providing input XP spectra are offered by the ''Input'' field (Fig.~\ref{fig:xpteal}):

\begin{enumerate}
    \item Choosing the option ''Use source\_ID dataset'' allows the user to upload an ascii file containing a list of {\gdr3} source identification number (source\_id), in a single column. The XP spectra for these sources, if available, are downloaded from the {\gaia} archive and processed. This option is preferable if the user already has a list of favourite sources available.
    \item Choosing the ''Use XP coefficients'' option allows the user to upload a file containing the XP continuous spectra as they can be downloaded via the datalink from the {\gaia} Archive, in csv format. This method is preferable if the user already has downloaded continuous XP spectra of the sources of interest from the {\gaia} Archive.
    \item Choosing the ''ADQL Query'' option allows the user to enter an ADQL query that is passed on to the {\gaia} Archive, and the available XP spectra for the sources matching the query are downloaded and processed.
    \item Choosing the ''Basic Query'' option opens a field in which simple conditions based on the fields in the DR3 {\gaia} source table can be combined to select sources. The available XP spectra for all matching sources are downloaded from the {\gaia} archive and processed.
\end{enumerate}

If the user wants the computation of EWs, the corresponding spectral lines need to be specified in the ''Options'' field of the web interface, under ''List of lines''.

There are currently two ways available to specify for which spectral lines to compute EWs. If ''Balmer lines'' is selected, the user can specify the hydrogen Balmer lines of interest. If ''File with list of wavelengths'' is chosen, the user can upload an ascii file containing the vacuum wavelengths, expressed in nm, of the spectral lines he is interested in.

Please note that in spectra of such low resolution as the {\Gaia} XP spectra, spectral lines can easily become blended. The consistent treatment of blended lines is not yet available. All lines are treated as being unblended, and the results for lines which are blended might be inaccurate.

For the computation of the EWs of lines different assumptions can be made. These assumptions can be chosen under ''Fitting of the lines''. If the option ''Narrow line'' is chosen, the line is treated as if it had a negligible intrinsic width as compared to the width of the XP LSF. This option is fastest in computation, the most robust approach and with the lowest random noise on the resulting EWs. Using this option for broad lines or absorption bands can however result in systematic errors, which depend on the kind of sources.

The computational stability can be improved by imposing a certain line profile. The user can choose a Gaussian profile, a Lorentzian profile, or a Voigt profile. For each of these line profiles, the user can optionally provide the Full Width Half Maximum (FWHM), in nm, of the profile. If the FWHM is not provided, the code determines this value internally, again based on the position of the inflexion points around a local extremum. As a rule, the more assumptions are made on the line profile or the intrinsic width of the line, the more stable the continuum interpolation becomes.

Finally, the user can choose to use truncation in the number of coefficients used. If ''Use truncated coefficients'' is chosen, only the first XP\_N\_RELEVANT\_BASES coefficients, as specified in the DR3 spectroscopic table, are used for the computation. This approach reduces noise on the spectra, but it might affect the representation of narrow spectral features such as spectral lines. It is therefore not recommended to use this option.

The web interface can provide different kinds of output. If lines for which to compute the EWs are specified, an output file with the results is provided. This file is a csv file that contains in its first column the {\gdr3} source\_id. Then, for each line five columns are provided. The line is either named by its Balmer designation, or by this wavelength in nm. The output data provided by the code are:
\begin{description}
    \item $W$: 
    EW of the line (nm), negative for absorption and positive for emission.
    \item {Werror:} Standard error of the EW (nm).
    \item $p$: $p$-value quantifying the significance of the local extremum (between 0 and 1); larger values indicate a lower probability that the extremum is due to noise.
    \item $D$: 
    Line width relative to the LSF width. Values close to 1 are consistent with a narrow line. This quantity is affected by some systematic errors, mainly resulting from an imperfect LSF model, and some slag should be permitted. However, values clearly smaller than 1 (say, around 0.5) might be an indicator that the local extremum is actually not a line, but a noise pattern. Values clearly larger than 1 (say, close to 2) might indicate that either the narrow line approximation is poor, or that the pattern is not a line at all, independent of the $p$-value.
    \item order:
    Indicator of the detection order: ``0'' for extrema found in the spectrum, ``2'' for extrema found in the second derivative, and ``3'' when no significant extremum is detected and only an upper limit is provided. In this last case, W is set to 0, there are no entries for p and D, and Werror contains the upper limit for W corresponding to a 1$\sigma$ uncertainty.
\end{description}

If the output option ''All extrema in the spectra'' is chosen, then two further output files are provided, containing all local extrema present in the spectrum and its second derivative, respectively. These are csv files, with a semicolon, '';'', as the secondary separator. They contain, for BP and for RP:
\begin{description}
    \item source\_id: The {\gdr3} source\_id
    \item extrema\_position\_xp: The position of the local extrema, in pseudo-wavelength units.
    \item extrema\_positionError\_xp: The errors of the position of the local extrema, in pseudo-wavelength units.
    \item extrema\_kind\_xp: The kind of extrema, either ''minimum'' or ''maximum''.
    \item extrema\_estimSignif\_xp: The $p$-values of the extrema.
    \item extrema\_estimWidth\_xp: The estimated width of the extrema, measured as the distance between the inflexion points surrounding the extremum, in pseudo-wavelength units.
    \item extrema\_estimWidthError\_xp: the standard error on extrema\_estimWidth\_xp, in pseudo-wavelength units.
\end{description}

These two files are rather large, and the interpretation may not be straightforward, so requesting them is recommended only if they are really needed by the user. If the output option ''sampled spectra'' is chosen, then furthermore a file containing plots of all spectra are provided.

\section{Online tables}
\label{sec:tables}
This paper provides several online tables via CDS. Here we show brief extracts of their content.

\begin{table*}[!htbp]
    \centering
\small
    \caption{Sample of the information in an online table with information of the clusters analysed in this work.}

\begin{tabular}{c|cccccccccc}
Cluster & RA & DEC & $E(\bprp)$ & $A_G$ & $A_V$ & $E(H-K)$ & $E(J-K)$ & $E(G_{\rm BP}-K)$ & GoodExcess & OC\\
\hline
1636-283&249.856&-28.399&2.473&5.820&4.722&0.380&0.626&5.416&false&false\\
ADS\_16795&352.592&58.553&0.081&0.183&0.168&0.008&0.013&0.147&false&true\\
AH03\_J0748+26.9&117.156&-26.973&0.554&1.260&1.128&0.079&0.130&1.204&true&true\\
.\\
.\\
.\\
         & 
    \end{tabular}
\tablefoot{The table contains some information directly extracted from \cite{Hunt2023} (right ascension, RA, declination, DEC and interstellar absorption in $V$ band, $A_V^{\rm Hunt}$) and others derived in this paper (the median colour of the cluster, $\langle\bprp\rangle$, the interstellar absorption in $G$ (using Eq.~\ref{eq:AGvsEBPRP_Basel}) and $V$ (using Eq.~\ref{eq:AVvsEBPRP_Basel}) bands, $A_G^{\rm pred}$, using Eq.~\ref{eq:AGvsEBPRP_Basel} and $\langle\bprp\rangle$ in this table, the colour excesses of the clusters derived using Eqs.~\ref{eq:EJHvsEBPRP_Basel}--\ref{eq:EBPKvsEBPRP}, and two flags indicating if we considered the cluster as an OC and if the colour excess derived can be considered of good quality, according to the filters mentioned in Sect.~\ref{sec:colourexcessOCs}).}
    \label{tab:clusters}
\end{table*}

\begin{table*}[!htbp]
    \centering
\caption{Sample of the information contained in an online table with the information of the members of the clusters analysed in this work.}
    \begin{tabular}{c|
    c
    cccccc
    c
    cc}
srcId & 
\ldots &
\Ha & $\sigma_{H\alpha}$ & $p_{H\alpha}$ & $D_{H\alpha}$ & order$_{H\alpha}$ & RelEmis$_{H\alpha}$ &
\ldots &
OC & Emitter\\
        \hline
5866370214722778880	& 
\ldots &
-0.237 &	0.094 & 0.931 & 0.898 & 2 & 0.140 &
\ldots &
T& F\\
5866193811838054016 & 
\ldots &
-0.2811 & 0.056 & 0.999 & 0.870 & 2 & 0.246 &
\ldots &
T & F \\
276804702298080768 & 
\ldots &
-0.137 & 0.077 & 0.531 & 2.033 & 2 & 0.278 &
\ldots &
T & F\\
.\\
.\\
.\\
         & 
    \end{tabular}
    \tablefoot{The members included here were extracted from \citealt{Hunt2023}. The table contains some information extracted from {\gdr3} (source identification, position, magnitude and colour). The rest of the columns are derived in this work. 
    For each source the width of the \Ha, \Hb and \Hg is provided (together with some relevant parameters of this determination). Finally two boolean flags indicating if a given star was considered to be part of an OC and if it is considered as an emitter in the previous literature are provided. }
    \label{tab:members}
\end{table*}

\begin{table}[!htbp]
    \centering
    \caption{Sample of the mastercurves of the de-reddened width of the Balmer EWs as a function of the \bprp colour.}
\begin{tabular}{c|ccc}
         \bprp & \Ha &\Hb & \Hg \\
         \hline
-0.5 & -0.240 & 0.562 & 0.171 \\
-0.499 & -0.241 & 0.558 & 0.170 \\
-0.498 & -0.242 & 0.554 & 0.168 \\
.\\
.\\
.\\
         & 
    \end{tabular}
    \label{tab:mastercurve}
\end{table}

\FloatBarrier 
\clearpage

\end{appendix}
\end{document}